\documentclass[aps,prb,reprint,a4paper,floatfix,groupedaddress,amsmath,amsfonts,amssymb]{revtex4-1}

\pdfoutput=1

\usepackage{mathptmx}
\usepackage[scaled=0.9]{helvet}
\usepackage[utf8]{inputenc} 
\usepackage{graphicx}
\usepackage[dvipsnames]{xcolor}
\usepackage[pdftex]{hyperref}
\hypersetup{
	pdftitle={Stacking faults as quantum wells in nanowires-Density of states oscillator strength and radiative efficiency},
	colorlinks,
	citecolor=blue
	}
\usepackage[textwidth=1.5cm]{todonotes}
\usepackage{textcomp}
\usepackage{sidecap}
\usepackage[
,textwidth=17.5cm
,textheight=23.5cm
,verbose
,dvips
]{geometry}
\usepackage{xspace}

\newcommand{\dox}{$(D^{0},X)$\xspace}
\newcommand{\sfx}{$(I_{1},X)$\xspace}

\begin{document}

\title{Stacking faults as quantum wells in nanowires:\\ Density of states, oscillator strength and radiative efficiency}

\author{P. Corfdir}
\email{corfdir@pdi-berlin.de}
\author{C. Hauswald}
\author{J. K. Zettler}
\author{T. Flissikowski}
\author{J. Lähnemann}
\author{S. Fernández-Garrido}
\author{L. Geelhaar}
\author{H. T. Grahn}
\author{O. Brandt}
\affiliation{Paul-Drude-Institut für Festkörperelektronik, Hausvogteiplatz 5–7, 10117 Berlin, Germany}

\date{\today}

\begin{abstract}
We investigate the nature of excitons bound to $I_1$ basal-plane stacking faults [$(I_{1},X)$] in GaN nanowire ensembles by continuous-wave and time-resolved photoluminescence spectroscopy. Based on the linear increase of the radiative lifetime of these excitons with temperature, they are demonstrated to exhibit a two-dimensional density of states, i.\,e., a basal-plane stacking fault acts as a quantum well. From the slope of the linear increase, we determine the oscillator strength of the \sfx and show that the value obtained reflects the presence of large internal electrostatic fields across the stacking fault.  While the recombination of donor-bound and free excitons in the GaN nanowire ensemble is dominated by nonradiative phenonema already at 10~K, we observe that the \sfx recombines purely radiatively up to 60~K. This finding provides important insight into the nonradiative recombination processes in GaN nanowires. First, the radiative lifetime of about 6~ns measured at 60~K sets an upper limit for the surface recombination velocity of 450~cm\,s$^{-1}$ considering the nanowires mean diameter of 105~nm. Second, the density of nonradiative centers responsible for the fast decay of donor-bound and free excitons cannot be higher than $2 \times 10^{16}$~cm$^{-3}$. As a consequence, the nonradiative decay of donor-bound excitons in these GaN nanowire ensembles has to occur indirectly via the free exciton state. 
\end{abstract}

\graphicspath{{figures/}}


\maketitle

\section{Introduction}
One remarkable property of III-V semiconductor nanowires is their ability to adopt both the wurtzite or zincblende crystal structure.\cite{Glas2007} This fact has opened the possibility to realize crystal-phase quantum structures, i.\,e., heterostructures in which confinement is achieved by alternation of the crystal structure of a given semiconductor.\cite{Caroff2008,Algra2008,Akopian2010,Corfdir2013} The major attractions of crystal-phase quantum structures are the absence of strain and alloy disorder as well as the atomically abrupt interfaces between the constituent crystal phases,\cite{Akopian2010,Bolinsson2011} i.\,e., they are expected to possess an exceptionally high structural perfection. In fact, very narrow linewidths have been reported for the characteristic emission lines observed for these structures.\cite{Akopian2010,Jacopin2011,Lahnemann2012,Graham2013} Consequently, crystal-phase quantum structures have been promoted as candidates  for optoelectronic applications such as lasers or single-photon emitters.\cite{Akopian2010,Castelletto2014,Kouno2014} Recent progress in the understanding of growth mechanisms has made it possible to grow nanowires with controlled crystal-phase superlattices \cite{Caroff2008,Algra2008} or nanowires with only a few crystal-phase quantum discs.\cite{Akopian2010,Vu2013,Ahtapodov2012} 

The prototypic crystal-phase quantum structure in a wurtzite crystal is the intrinsic $I_1$ basal-plane stacking fault (BSF). It has been proposed that the $I_1$ BSF may be considered to represent a three monolayer thick zincblende quantum well (QW) in a wurtzite matrix.\cite{Albrecht1997} This assertion was supported theoretically by the ab-initio calculations in Ref.~\onlinecite{Stampfl1998} and has recently been confirmed experimentally in Ref.~\onlinecite{Korona2014} for $I_1$ BSFs in GaN. However, little is known about the nature of the excitons confined in these QWs. In particular, a quantitative knowledge of their oscillator strength and radiative efficiency would be essential for a qualified assessment of the actual potential of crystal-phase quantum structures for optoelectronic applications. 

In this work, we investigate the dynamics of excitons bound to $I_1$ BSFs in GaN nanowire ensembles [$(I_{1},X)$]. Based on the evolution of their radiative lifetime with temperature, we show that the \sfx exhibit a two-dimensional density of states, and we determine their oscillator strength. Their decay is purely radiative up to 60~K, allowing us to estimate the maximum surface recombination velocity of the nanowires \emph{M}-plane sidewalls as well as the maximum density of nonradiative centers in the nanowires. For higher temperatures, the radiative efficiency of the \sfx decreases as a result of their thermal escape from the BSFs. 

\section{Experimental}

The GaN nanowire ensemble studied here is formed spontaneously during plasma-assisted molecular beam epitaxy on a Si(111) substrate.\cite{Garrido2009,Geelhaar2011,Consonni2013} The nanowires are on the average 2~\textmu m long and exhibit a mean diameter of 105~nm. The sample was prepared at high substrate temperature (865 °C) and a Ga/N flux ratio higher than one to compensate for the high Ga desorption rate at this temperature.\cite{Heying2000} As a result of these conditions, we observe melt-back etching of the Si substrate arising from the formation of a Ga-Si eutectic alloy.\cite{Corfdir2014} Photoluminescence (PL) experiments revealed that the melt back-etching of the Si substrate during the formation of the GaN nanowire ensemble leads to both unintentional Si doping and the formation of a high density of BSFs.\cite{Corfdir2014} The latter phenomenon is rarely observed in GaN nanowires prepared using conventional growth conditions, i.\,e.,  substrate temperatures $\leq 820$\,°C and Ga/N flux ratios well below one.\cite{Garrido2009,Corfdir2009c} More details about the substrate preparation and the specific growth conditions can be found elsewhere.\cite{Corfdir2014}

The present sample was chosen for its narrow near-band-edge and strong BSF emission at 10~K, both of which are characteristics for GaN nanowire ensembles grown above 850\,°C.\cite{Corfdir2014} However, the conclusions drawn in this article have been verified for several samples with very different morphologies and BSF densities (see supplementary information).

\begin{figure}
\includegraphics[scale=1]{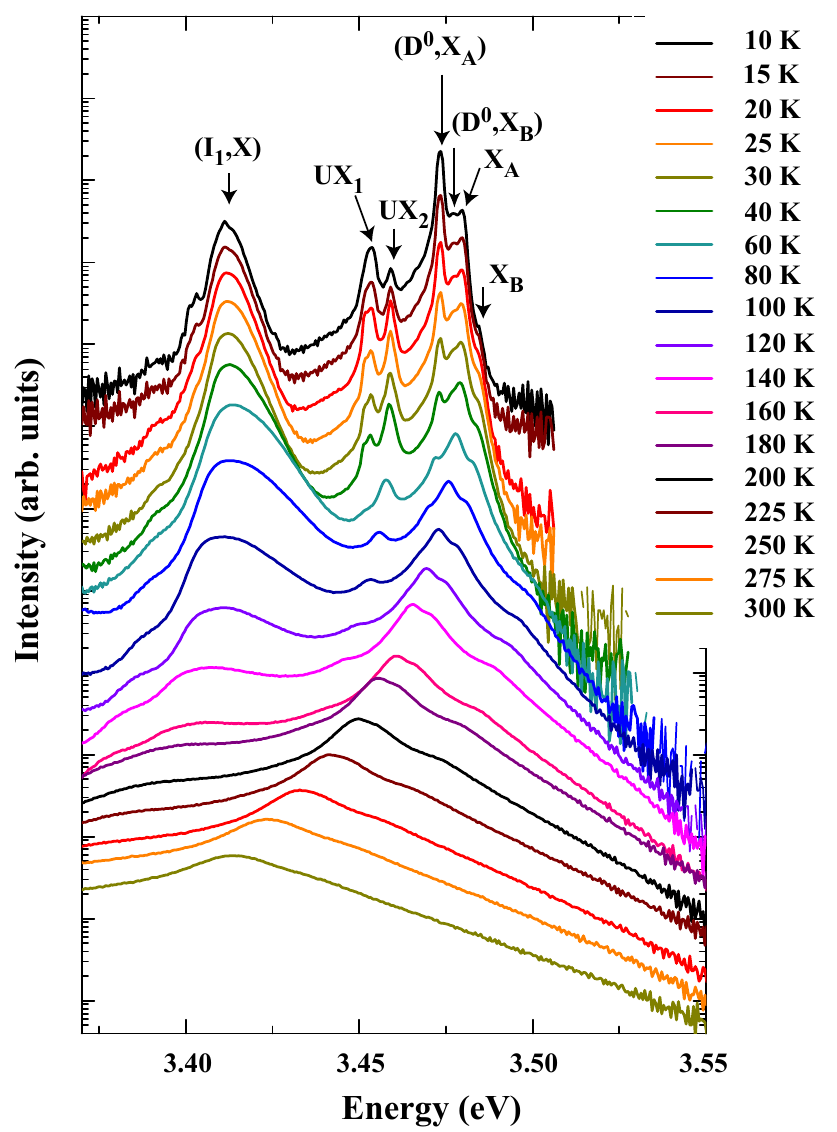}
\caption {(color online): Continuous-wave PL spectra of the GaN nanowire ensemble under consideration for various temperatures as indicated in the figure. The spectra are shifted vertically for clarity. The major transitions are labeled according to their origin.}
\label{fig:Figure1}
\end{figure}

Continuous-wave (cw) PL experiments were carried out with a HeCd laser for excitation ($\lambda = 325$~nm).  The laser beam with a power of less than 150~nW was focused onto the sample surface to a 1~\textmu m diameter spot using a microscope objective with a numerical aperture of 0.65. The emitted light was collected using the same objective and then directed to a monochromator (focal length 80~cm, 600 lines per mm grating) followed by a charge-coupled (CCD) device camera. Time-resolved (TR) PL spectroscopy was performed using the second harmonic ($\lambda = 325$~nm) of fs pulses obtained from an optical parametric oscillator pumped by a Ti:Sapphire laser with a repetition rate of 76~MHz. We used an energy fluence per pulse of 1~\textmu J\,cm$^{-2}$. The emitted light was dispersed using a monochromator with a 22~cm focal length and a 1800 lines per mm grating. The dispersed light was then directed to a streak camera that can be operated either in synchroscan or in single sweep mode. For both cw and TR~PL, the samples were mounted on a coldfinger cryostat whose temperature $T$ was varied between 10 and 300 K. For the TR~PL experiments, one expects a significant deviation between the lattice temperature and the effective temperature of the exciton gas. Consequently, throughout this paper, the temperature specified for TR~PL experiments is the effective carrier temperature, which we extract from the PL spectra as detailed in Ref.~\onlinecite{Corfdir2009c}. Cathodoluminescence spectroscopy (CL) was performed in a scanning electron microscope with the acceleration voltage and probe current set to 3~kV and 500~pA, respectively. The emitted light was collected with a parabolic mirror, dispersed by a monochromator (focal length 30~cm, 1200 lines per mm grating) and detected by a photomultiplier.  

\section{Results and discussion}

\subsection{Temperature-dependent spectra of the stacking-fault exciton}

\begin{SCfigure*}
\includegraphics[scale=1]{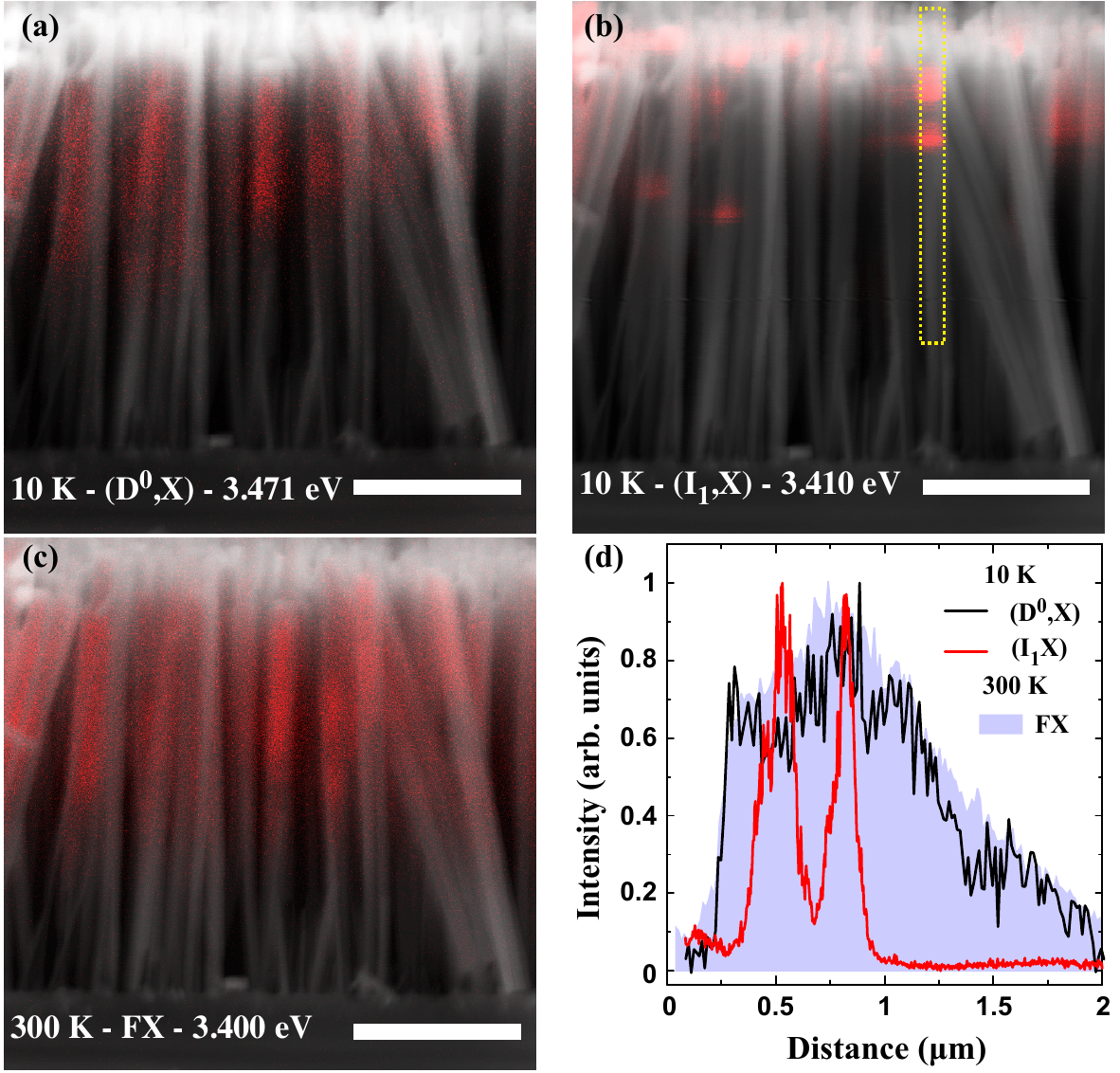}
\caption {(color online) Superposition of scanning electron micrographs (greyscale) and monochromatic CL maps (false color) taken at 10~K at the emission energy of (a) the \dox and (b) the \sfx. The scale bar corresponds to 1~\textmu m. (c) Same as (a) and (b) but taken at 300~K at the FX emission energy. (d) CL intensity of the \dox and \sfx transitions at 10~K and of the FX line at 300~K along the nanowire highlighted by the yellow rectangle in (b).}
\label{fig:Figure2}
\end{SCfigure*}

Figure \ref{fig:Figure1} shows the PL spectra of the GaN nanowire ensemble under investigation at temperatures between 10 and 300~K. At 10~K, the spectrum is dominated by a strong line at 3.472~eV that originates from the recombination of A excitons bound to neutral donors [$(D^{0},X_\text{A})$]. We attribute the weaker line at 3.476~eV to the emission from donor-bound B excitons [$(D^{0},X_\text{B})$]. The emission from free A excitons ($X_\text{A}$) and free B excitons ($X_\text{B}$) are also resolved at 3.4785 and 3.484~eV, respectively. In this paper, we are not interested in the detailed interplay between $(D^{0},X_\text{A})$ and $(D^{0},X_\text{B})$. Therefore, in the following, the emission intensity spectrally integrated over the $(D^{0},X_\text{A})$ and $(D^{0},X_\text{B})$ lines is simply denoted as \dox. Similarly, from now on the emission integrated over the $X_\text{A}$ and $X_\text{B}$ lines is labeled FX. On the low energy side of the \dox emission, we observe two bands centered at 3.452 and 3.458~eV labeled UX$_1$ and UX$_2$, respectively.\cite{Calleja2000} The origin of these bands has not been established beyond doubt, but we believe them to be related to the nanowire surface.\cite{Corfdir2009c,Lefebvre2011} Note that these transitions are superimposed onto those associated with the two-electron satellites of the \dox transitions from the core of the nanowires.\cite{Corfdir2009c} Finally, the broad band centered at 3.410~eV arises from the recombination of $(I_{1},X)$.\cite{Paskov2005,Guhne2008,Corfdir2009a,Lahnemann2012} We attribute the slight deviation of the BSF peak energy from values reported for BSFs in GaN layers\cite{Paskov2005,Guhne2008,Corfdir2009a} to variations in the local stacking fault density\cite{Corfdir2012} and in the magnitude of the built-in electric fields along the \emph{c}-axis.\cite{Corfdir2012,Lahnemann2012}

Low-temperature CL maps taken at the \dox and \sfx emission energies are displayed in Figs.~\ref{fig:Figure2}(a) and \ref{fig:Figure2}(b), respectively. While the first 1~\textmu m of the nanowires show a weak \dox CL, the \dox emission for the top part of the nanowire is intense [Fig.~\ref{fig:Figure2}(a)]. We attribute this observation to the fact that the density of donors and therefore the density of donor-bound excitons are larger at the top of the nanowires. It suggests that for our specific growth conditions, the increase in Si incorporation that results from the melt-back etching of the Si substrate does not take place from the beginning of the nanowire growth but occurs rather after the growth of about 1~\textmu m of material. Accordingly, as an increase in Si incorporation is accompanied by a reduction in BSF formation energy,\cite{Chisholm2000} the CL of \sfx is observed only for the top part of the nanowires [Fig.~\ref{fig:Figure2}(a)]. We emphasize that the intensities of the \dox and \sfx transitions are not spatially anti-correlated [Fig.~\ref{fig:Figure2}(d)]. Similar observation was made for layers with a BSF density on the order of 10$^4$~cm$^{-1}$,\cite{Corfdir2009a} a density that would correspond to the presence of only a few BSFs per nanowire. The absence of spatial anti-correlation between the \sfx and the \dox CL arises from the fact that the characteristic times for the recombination of FX and \dox and for the capture of excitons by the BSFs are on the same order, in agreement with the results of Ref. \onlinecite{Corfdir2009b}. There is also no anti-correlation between the intensities of the \sfx transition at 10~K and of the FX transition at 300~K [cf.\ Figs. \ref{fig:Figure2}(b)--\ref{fig:Figure2}(d)]. Since the diffusion length in GaN is much shorter than the nanowire length,\cite{Corfdir2011,Nogues2014} this observation indicates that BSFs in nanowires do not open an efficient nonradiative recombination channel. Finally, the FX CL at 300~K is stronger at the top than at the bottom of the nanowires, which we tentatively attribute to the permanent exchange occuring at 300~K between FX and \sfx.\cite{Corfdir2009a}

\begin{figure}
\includegraphics[scale=1]{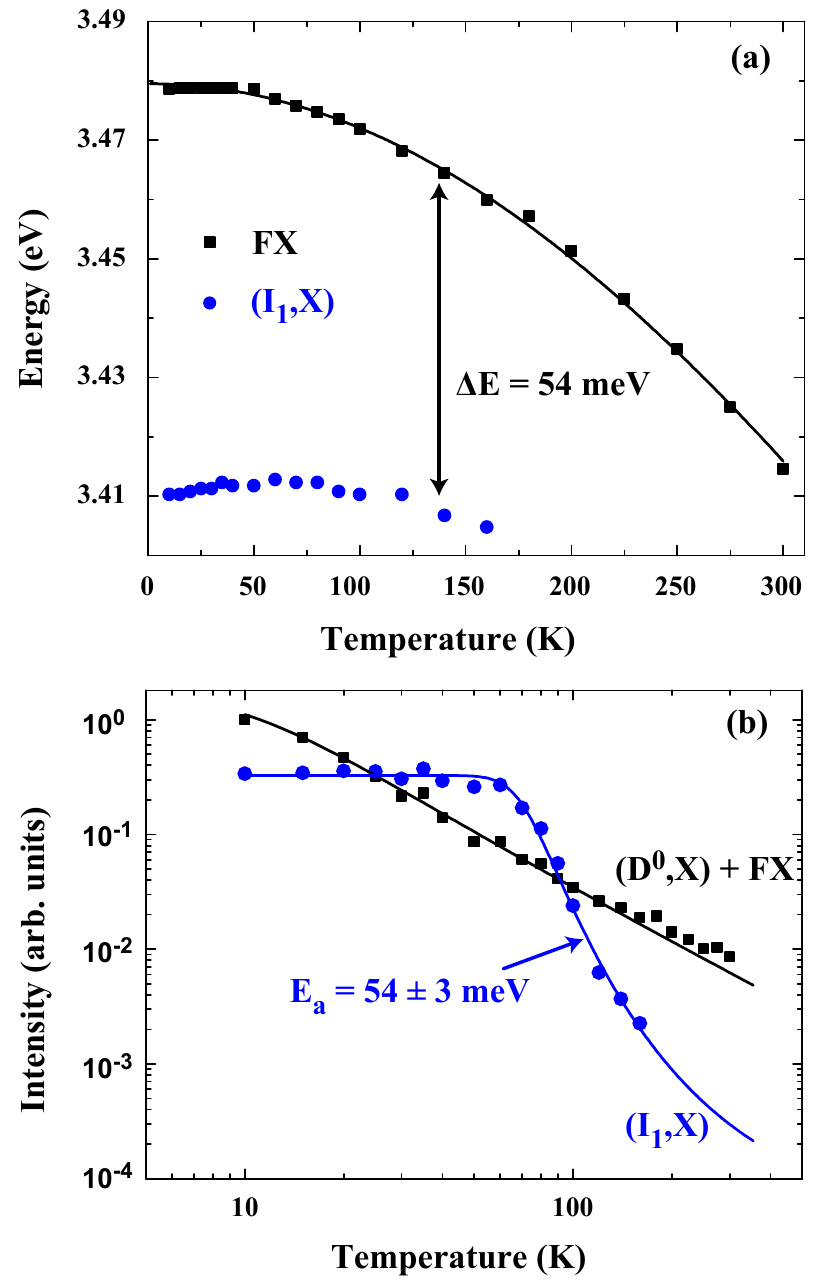}
\caption {(color online): (a) FX (squares) and \sfx (circles) emission energies as a function of $T$. The solid line is a fit to the free exciton emission energy using the expression derived by Pässler.\cite{Paessler2001} $\Delta E = 54$~meV is the energy difference between the FX and the delocalized \sfx. (b) Sum of the FX and \dox emission intensities (squares) and \sfx emission intensity (circles) as a function of $T$, displayed using a double logarithmic scale. The black solid line is a fit to the total FX and \dox intensity assuming that the nonradiative lifetime of FX is a constant and that its radiative lifetime increases with $T^{3/2}$. The blue solid line is the result of an Arrhenius fit to the \sfx emission intensity accounting for a nonradiative recombination channel with an activation energy $E_\text{A} = 54 \pm 3$~meV.}
\label{fig:Figure3}
\end{figure}

With increasing temperature, the dissociation of the \dox leads to the quenching of the corresponding transition to the benefit of the FX line (Fig.~\ref{fig:Figure1}). Between 10 and 300~K, the FX transition redshifts, following a temperature dependence well described by the expression derived by Pässler\cite{Paessler2001} with the parameters recommended for GaN(0001) layers as shown in Fig.~\ref{fig:Figure3}(a). In contrast to the FX transition, the energy of the \sfx band follows an half \textsf{S}-shaped temperature dependence: it blueshifts between 10 and 60~K and redshifts for larger $T$. This behavior reflects exciton localization along the BSF plane.\cite{Paskov2005,Corfdir2009a} Since BSFs are free of strain and alloy disorder and exhibit abrupt interfaces, intra-BSF localization has been attributed previously to the presence of neutral donors in the vicinity of the BSF plane.\cite{Corfdir2009a,Graham2013} For $T > 60$~K, the \sfx and FX transitions redshift and exhibit an energy difference of $\Delta E = 54$~meV.

Figure \ref{fig:Figure3}(b) shows the temperature dependence of the PL intensities of the \sfx transition and of the sum of the intensities of the \dox and FX transitions, obtained from a deconvolution of the spectra shown in Fig. \ref{fig:Figure1}. The latter decreases following a $T^{-3/2}$ dependence, demonstrating (i) that the dynamics of the FX and the \dox are dominated by nonradiative recombination already at cryogenic $T$, and (ii) that the decay rate of this nonradiative recombination channel is essentially constant over the whole temperature range since the radiative decay rate of the FX decreases with $T^{3/2}$.\cite{Hauswald2014b} In contrast, the  intensity of the \sfx transition remains constant between 10 and 60~K and quenches for larger $T$ following an Arrhenius behavior with an activation energy $E_\text{a} = 54 \pm 3$~meV. This value is equal to the energy difference $\Delta E$ determined from the data shown in Fig.~\ref{fig:Figure3}(a), and we therefore attribute the quenching of the \sfx emission at $T > 60$~K to the thermally activated detrapping of excitons as also found for GaN layers\cite{Paskov2005} and GaAs nanowires.\cite{Graham2013,Rudolph2013} For nonpolar GaN layers,\cite{Paskov2005} the delocalization of \sfx between 10 and 50~K is accompanied by the activation of a nonradiative recombination channel. This channel has been ascribed to the capture of \sfx by the nonradiative partial dislocations, which close the BSF plane.\cite{Guhne2008,Wu2008} Since BSFs in nanowires extend across the whole nanowire section,\cite{Algra2008,Corfdir2013} this additional nonradiative process is not observed in our sample [cf.\ Fig.~\ref{fig:Figure3}(b)].

The constant \sfx emission intensity between 10 and 60~K may reflect two distinct scenarios. The first obvious explanation for this observation is that the \sfx decay is purely radiative over this temperature range. However, since the nonradiative lifetime of the FX and the \dox is constant between 10 and 300~K, one could argue that it could also be constant for the \sfx. Observing a constant \sfx emission intensity between 10 and 60~K would simply require the \sfx radiative lifetime to be constant as well, irrespective of the overall internal quantum efficiency of \sfx. This second scenario would take place if \sfx are localized between 10 and 60~K; the half \textsf{S}-shape dependence observed in Fig.~\ref{fig:Figure3}(a) would have to arise from the redistribution of excitons among these localized states. In the following, we carry out TR~PL on our GaN nanowire ensemble focusing on the dynamics of the \sfx state to distinguish between the two scenarios discussed above.

\subsection{Density of states and oscillator strength of the stacking-fault exciton}

\begin{figure*}
\includegraphics[scale=1]{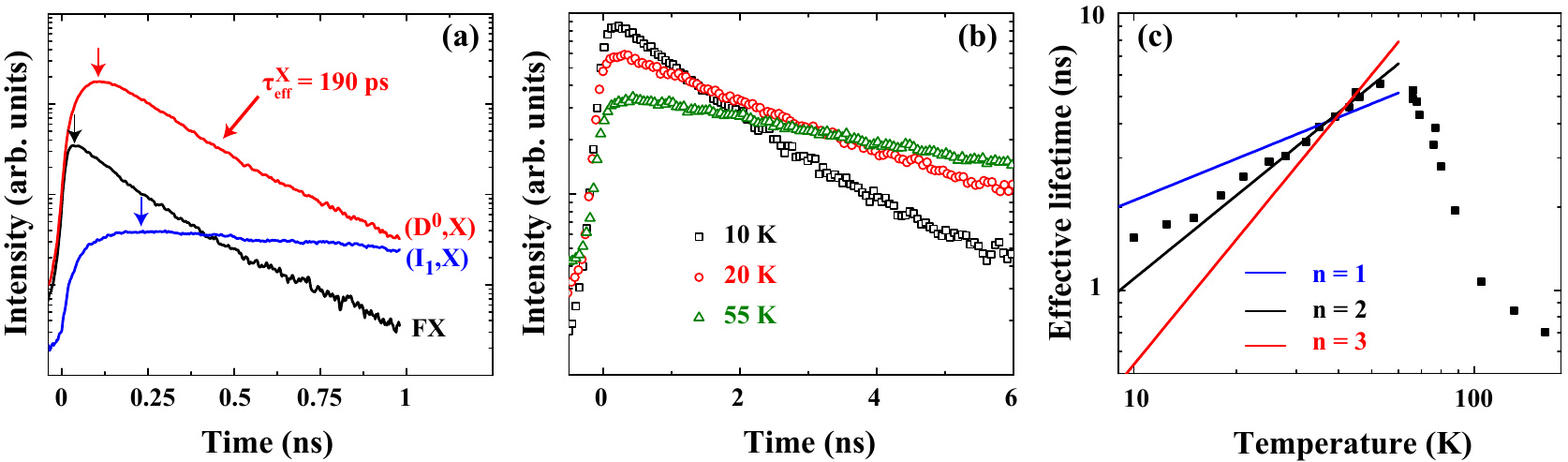}
\caption {(color online): (a) PL intensity transients of the FX, \dox, and \sfx transitions at 10~K. The temporal resolution was set to 20~ps. The vertical arrows indicate the time at which each transient reaches its maximum intensity. The effective lifetime $\tau_\text{eff}^\text{X} = 190$~ps has been obtained from an exponential fit to the \dox transient. (b) PL intensity transients of the \sfx transition at 10, 20, and 55~K. The temporal resolution was set to 160~ps. An exponential fit of these transients yields $\tau_\text{eff}^\text{BSF} = $1.6, 2.9, and 5.6~ns for $T = $10, 20 and 55~K, respectively. (c) Temperature dependence of $\tau_\text{eff}^\text{BSF}$ (squares). The solid lines show the result of the fits using Eq.~(\ref{eqn:fit}) for $n = 1$ ($A = 0.666$~ns\,K$^{-1/2}$), $n = 2$ ( $A = 0.110$~ns\,K$^{-1}$), and $n = 3$ ($A = 0.017$~ns\,K$^{-3/2}$).}
\label{fig:Figure4}
\end{figure*}

PL transients recorded at 10~K of the FX, \dox and \sfx transitions are shown in Fig.~\ref{fig:Figure4}(a). While the FX emission intensity increases almost instantaneously, the \dox emission intensity reaches its maximum after a delay of about $70$~ps, which corresponds to the time required to capture excitons by neutral donors, in agreement with values from previous reports.\cite{Corfdir2009c,Hauswald2013} For longer time delays, the FX and \dox transitions decay in parallel, indicating the thermalization of these two distributions of excitons.\cite{Corfdir2009c,Hauswald2013} The \dox and FX transients are nearly exponential over the first 0.6~ns of the decay. They exhibit an effective exciton lifetime $\tau_\text{eff}^\text{X}=190$~ps, which is given by 

\begin{align}
\label{eqn:taueff}
\frac{1}{\tau_\text{eff}^\text{X}} = \frac{1}{\tau_\text{r}} + \frac{1}{\tau_\text{nr}},
\end{align}
where $\tau_\text{r}$ and $\tau_\text{nr}$ are the exciton radiative and nonradiative lifetimes, respectively. In high-quality GaN layers, $\tau_\text{r}$ for the \dox state is at least 1~ns.\cite{Monemar2010} Therefore, the decay of the thermalized distribution of the FX and the \dox in our GaN nanowires is certainly dominated by nonradiative phenomena.\cite{Hauswald2014b} After about 0.6~ns, the FX and \dox transients deviate from a pure exponential behavior due to the coupling between these states and deeper states such as acceptor-bound excitons or the UX states.\cite{Hauswald2013}

Figure \ref{fig:Figure4}(a) shows that both the increase and the decay of the \sfx intensity are much slower than those measured for the \dox intensity. First, the \sfx transition reaches its maximum intensity 125~ps after the \dox line. The factor limiting the increase of the \sfx emission intensity is the transport of excitons from nanowire segments free of extended defects to the BSFs. This transport is inefficient as it takes place at low temperatures via the hopping of excitons from one donor to the next.\cite{Corfdir2009b} Second, the effective decay time for the \sfx recombination is $\tau_\text{eff}^\text{BSF} = 1.6$~ns, a decay time significantly longer than that reported so far for BSFs in group-III-nitride layers.\cite{Lahourcade2008,Corfdir2009a,Corfdir2009b,Badcock2012,Kagaya2011,Furusawa2013} Moreover, this decay time is almost ten times longer than $\tau_\text{eff}^\text{X}$, which suggests that the \sfx are much less affected by nonradiative recombination than both the \dox and FX. 

In fact, Figs.~\ref{fig:Figure4}(b) and \ref{fig:Figure4}(c) show that $\tau_\text{eff}^\text{BSF}$ increases from 1.6 to 5.6~ns between 10 and 60~K, and decreases subsequently to 0.7~ns when $T$ is increased further from 60 to 160~K. As detailed in Ref.~\onlinecite{Corfdir2011}, measuring simultaneously a constant emission intensity and an increase in $\tau_\text{eff}^\text{BSF}$ demonstrates the dominance of radiative phenomena over nonradiative ones. In other words, between 5 and 60~K, $\tau_\text{r}^\text{BSF} \approx \tau_\text{eff}^\text{BSF}$. For larger $T$, the thermal escape of charge carriers from the BSFs to the wurtzite matrix is activated [cf.\ Fig.~\ref{fig:Figure3}(b)]. Considering that the density of states in the GaN wurtzite matrix is certainly much higher than in the BSFs, the dynamics of the \sfx is then governed by that of the FX.\cite{Corfdir2012b} Since the FX decay is mostly nonradiative already at 10~K, the thermal detrapping of excitons from BSFs for $T > 60$~K is accompanied by a rapid decrease in $\tau_\text{eff}^\text{BSF}$.

The ability to directly measure the increase of $\tau_\text{r}^\text{BSF}$ with $T$ facilitates the investigation of fundamental electronic properties of BSFs. The radiative lifetime $\tau_\text{r}$ for a localized exciton is proportional to the exciton coherence volume and therefore does not vary with $T$.\cite{Rashba1962} In contrast, the radiative lifetime of free excitons in the bulk (provided that the exciton scattering rates are faster than the exciton-photon coupling rate) or confined in a quantum well or quantum wire increases with $T$.\cite{Hooft1987,Andreani1991,Citrin1993a} This increase results from the fact that for higher $T$ excitons populate states that cannot couple to light. Using the results of Refs.~\onlinecite{Rashba1962,Hooft1987,Andreani1991,Citrin1993a}, $\tau_r$ is given by

\begin{align}
\label{eqn:fit}
\tau_\text{r} = A T^{n/2} ,
\end{align}
where $0\leq n \leq 3$ is the dimensionality of the system and $A$ is a constant. For $T> 20$~K, the increase in $\tau_\text{r}^\text{BSF}$ is linear ($n = 2$), and we obtain $A = 0.11$~ns\,K$^{-1}$. Assuming $n=1$ or $n=3$ does not lead to an acceptable fit of the data [cf.\ Fig.~\ref{fig:Figure4}(c)]. This finding shows conclusively that the density of states for free excitons confined to a BSF is two-dimensional. Consequently, we experimentally confirm the proposal by \citet{Albrecht1997} that a BSF forms a QW. We recall that we investigate here the emission properties of BSFs in nanowires with a mean diameter of 105~nm, which is about 30 times larger than the exciton Bohr radius.\cite{Corfdir2012} When contained in planar heterostructures with the growth axis perpendicular to the BSF plane or in thinner nanowires, BSFs may lead to the formation of quantum structures with lower dimensionality such as quantum wires\cite{Dussaigne2011} or quantum dots.\cite{Akopian2010}

The deviation of $\tau_\text{r}^\text{BSF}$ from a linear behavior for $T < 20$~K signifies the presence of localization centers in the vicinity of the BSF plane. Assuming that the \sfx relaxes to localization centers with a density $N_\text{D}$ within a BSF faster than it decays, we obtain\cite{Citrin1993,Corfdir2011}

\begin{align}
\label{eqn:taur}
\tau_\text{r}^\text{BSF}=\frac {N_\text{loc}+N_\text{fr}}{N_\text{loc}\Gamma_\text{loc}+N_\text{fr}\Gamma_\text{fr}},
\end{align}
where $N_\text{loc}$ and $N_\text{fr}$ are the densities of localized and free \sfx states, respectively, with decay rates $\Gamma_\text{loc}$ and $\Gamma_\text{fr}$, respectively. We assume that at 10~K all \sfx are localized and exhibit a radiative lifetime of $\tau_\text{loc} = \Gamma_\text{loc}^{-1}$ = 1.6~ns [cf.\ Fig.~\ref{fig:Figure4}(b)]. It follows directly from Eqs.~(\ref{eqn:fit}) and (\ref{eqn:taur}) that $\tau_\text{r}^\text{BSF}$ is a constant for $N_\text{fr} \ll N_\text{loc}$, but a linear function of $T$ for $N_\text{fr} \gg N_\text{loc}$. Interestingly, the slope $A$ for the increase of $\tau_\text{r}^\text{BSF}$ with $T$ is affected by the presence of potential fluctations even when $N_\text{fr} \gg N_\text{loc}$. Specifically, $A$ is found to be inversely proportional to the density of localized states $N_\text{D}$ for sufficiently high densities (see Appendix\ \ref{AppendixA} for a derivation of this result):

\begin{align}
\label{eqn:slope}
 A = 2 \gamma k_B \left(f_\text{osc} + \frac{ \gamma  \pi \hbar^2 N_\text{D} }{ M \tau_\text{loc}}\right)^{-1},
\end{align}
where $k_B$ is the Boltzmann constant, $f_\text{osc}$ the  oscillator strength of the free \sfx, $N_\text{D}$ the areal density of localized states in the BSF, and $M = 1.2m_0$ the exciton translational mass. The coefficient $\gamma$ is given by:   

\begin{align}
\label{eqn:gamma}
 \gamma = \frac{3 m_0 c \epsilon_0 \sqrt{\epsilon}}{E_1e^2} ,
\end{align}
where $c$ is the speed of light in vacuum, $\epsilon = 9.5$ the relative permittivity of GaN, $e$ the elementary charge, and $E_1$ the exciton kinetic energy above which excitons cannot couple to light. In GaN, $E_1$ is on the order of 0.1~meV.\cite{Corfdir2011}

The value of $A$ measured for the \sfx state is much larger than for excitons in nonpolar (Al,Ga)N/GaN QWs, for which $A$ amounts to only a few ps\,K$^{-1}$.\cite{Corfdir2012b,Rosales2014} This result confirms that crystal-phase quantum structures exhibit improved interfaces and reduced densities of localization centers as compared to planar heterostructures. Quantitatively, we can estimate an upper limit for the oscillator strength $f_\text{osc} < 1.5 \times 10^{12}$~cm$^{-2}$ for $N_\text{D} \rightarrow 0$ [Eq.~(\ref{eqn:slope})]. This  value is two orders of magnitude smaller than the value expected for free excitons in (Al,Ga)N/GaN QWs free of electrostatic fields.\cite{Corfdir2011} The small value of $f_\text{osc}$ obtained here is caused by the presence of built-in electric fields that spatially separate the electron and hole wavefunctions.\cite{Corfdir2012} These fields arise from the discontinuity of the spontaneous polarization field at the interface between zincblende and wurtzite GaN and can exhibit a magnitude of a few MV/cm.\cite{Sun2002,Lahnemann2012} Note that we have obtained similar values for $N_\text{D}$, $f_\text{osc}$, and $\tau_\text{eff}^\text{BSF}$ at 10~K for all the nanowire samples investigated in the course of this work (see the supplemental material). 

Our values differ from the results of \citet{Korona2014}, who have recently reported a radiative lifetime of $\tau_\text{r}^\text{BSF} \approx 0.75$~ns at 10~K and a slope of $A = 0.015$~ns\,K$^{-1}$. This discrepancy may simply arise due to the presence of an additional nonradiative channel enhancing the recombination rate at higher $T$. Alternatively, it may be caused by the different type of samples under investigation. Indeed, \citet{Korona2014} studied an intentionally coalesced nanowire ensemble with a very high density of BSFs, as reflected by the fact that the \sfx intensity in their spectra is orders of magnitude higher than that of any other transition. For samples with such a high density of BSFs, adjacent BSFs will couple electronically.\cite{Paskov2005} As shown in Ref.~\onlinecite{Corfdir2012}, this coupling between BSFs may lead to an increase in $f_\text{osc}$, resulting in a decrease of both $\tau_\text{r}^\text{BSF}$ and $A$ with respect to the values obtained here for nanowire ensembles with a low BSF density.

The crossover between dominantly radiative and nonradiative recombination occurs at 60~K for the sample under investigation. At this temperature, the nonradiative lifetime ($\tau_\text{nr}^\text{BSF}$) of\sfx is necessarily larger than $\tau_\text{eff}^\text{BSF} \approx 6$~ns. At the same time, the \sfx is delocalized at this temperature and is free to diffuse along the BSF plane. Since the nanowire diameter $\phi$ is comparable to the diffusion length of the free exciton,\cite{Nogues2014} the \sfx will probe the \emph{M}-plane nanowire sidewall facets and recombine there nonradiatively via surface states if such states are present. If we assume that the quenching of the \sfx emission intensity arises exclusively from surface recombination, the surface recombination velocity is $S = \phi / (4 \tau_\text{nr}^\text{BSF}) < 450$~cm\,s$^{-1}$, where $\phi = 105$~nm is the mean nanowire diameter. This value of $S$ is more than an order of magnitude smaller than what has been reported in Refs.~\onlinecite{Gorgis2012,Schlager2008} for GaN nanowires at 10 and 300~K, respectively. The surface recombination in GaN nanowires is thus either strongly temperature dependent, or the nonradiative recombination observed in Refs.~\onlinecite{Gorgis2012,Schlager2008} does not arise solely from the nanowire surface. However, since the nonradiative lifetime of FX and (D$^0$,X) has been found almost constant between 10 and 300~K,\cite{Hauswald2014b} the nanowire surface is most probably not the main nonradiative decay channel for excitons in our GaN nanowires.

\section{Conclusions}

In conclusion, we have observed with time-resolved PL experiments on GaN nanowire ensembles that the density of states of excitons bound to $I_1$ BSFs is two-dimensional: this finding demonstrates that BSFs form quantum wells. The recombination of the \sfx is purely radiative up to 60~K. The effective lifetime measured for \sfx at 60~K indicates that the recombination velocity in our GaN nanowires is smaller than 450~cm\,s$^{-1}$. For larger $T$, the \sfx escape thermally from the BSFs and recombine nonradiatively, leading to a decrease in effective decay time for the \sfx recombination.

The fast decay of the FX and the \dox at 10~K in GaN nanowires is dominated by a nonradiative channel that is most likely opened by point defects.\cite{Hauswald2014a, Hauswald2014b} To explain the fast decay of the majority of \dox states, the density of nonradiative centers has to be higher than $5 \times 10^{17}$~cm$^{-3}$.\cite{Hauswald2014a} However, the FX and \dox states were found to efficiently couple in GaN nanowires even at low temperatures [cf.\ Fig.~\ref{fig:Figure4}(a)], and the \dox may consequently decay \emph{indirectly} via the FX.\cite{Hauswald2014b} In this case, it is not in general possible to ascertain whether it is the \dox or the FX state, which is actually affected by the nonradiative recombination.\cite{Hauswald2014b}

Using the BSFs as a probe, we are able to distinguish between these two scenarios. A density of nonradiative centers of $5 \times 10^{17}$~cm$^{-3}$ is inconsistent with the observation of a dominantly radiative recombination in BSFs. As we probe several hundreds of BSFs in our experiments, the majority of them are required to be free of these nonradiative centers. Hence, the density of these centers must be at least significantly lower than the inverse volume experienced by an \sfx. Since the extent of the \sfx wavefunction along the nanowire axis amounts to about 6~nm\cite{Corfdir2012} and the radial one is equal to the nanowire diameter of 105~nm, the density of nonradiative centers \emph{must} be well below $2 \times 10^{16}$~cm$^{-3}$. Evidently, this density is not compatible with the one required for the direct nonradiative decay of the \dox. If, however, the \dox decays indirectly via the FX, even densities well below $1 \times 10^{16}$~cm$^{-3}$ are sufficient for opening an efficient nonradiative channel, since the diffusion length of the FX is reported to be larger than 50~nm.\cite{Nogues2014} 

\acknowledgements

The authors would like to thank Uwe Jahn for a critical reading of the manuscript and Henning Riechert for continuous encouragement and support. Partial financial support of this work by the Deutsche Forschungsgemeinschaft within SFB 951 is gratefully acknowledged.

\appendix

\section{Radiative lifetime of the stacking-fault exciton}
\label{AppendixA}

For evaluating the temperature ($T$) dependence of the radiative lifetime of excitons confined in an $I_{1}$ BSF in the presence of potential fluctuations, we follow the procedure established by \citet{Citrin1993}. The areal density of localized states in the BSF is denoted by $N_\text{D}$. The distributions of localized ($N_\text{loc}$) and free excitons ($N_\text{fr}$) with radiative decay rates $\Gamma_\text{loc}$ and $\Gamma_\text{fr}$, respectively, are supposed to be in thermal equilibrium. The radiative decay rate $\Gamma_\text{r}^\text{BSF}$ for the thermalized population of excitons confined to the BSF is then

\begin{align}
\label{eqn:equationA1}
\Gamma_\text{r}^\text{BSF}=\frac {N_\text{loc}\Gamma_\text{loc}+N_\text{fr}\Gamma_\text{fr}}{N_\text{loc}+N_\text{fr}}.
\end{align}
When $T$ is high enough for $N_\text{loc} \ll N_\text{fr}$, we obtain

\begin{align}
\label{eqn:equationA2}
\tau_\text{r}^\text{BSF} = 1/\Gamma_\text{r}^\text{BSF}   =\frac{1}{\Gamma_\text{fr}} \left( 1 + \frac{N_\text{loc}\Gamma_\text{loc}}{N_\text{fr} \Gamma_\text{fr} } \right)^{-1}.
\end{align}
If $N_\text{loc}$ and $N_\text{fr}$ follow Boltzmann distributions, the ratio between localized and free exciton densities is given by

\begin{align}
\label{eqn:equationA3}
\frac{N_\text{loc}}{N_\text{fr}} = \frac{N_\text{D}\pi\hbar^2}{2Mk_BT}\exp\left({\frac{E_\text{loc}}{k_BT}}\right),
\end{align}
where $M$ is the exciton translational mass and $E_\text{loc}$ is the exciton localization energy. While $\Gamma_\text{loc}$ is independent of $T$,\cite{Rashba1962} $\Gamma_\text{fr}$ decreases linearly with $T$:\cite{Andreani1991}

\begin{align}
\label{eqn:equationA4}
\Gamma_\text{fr} = \frac{2E_1}{3k_BT} \frac{e^2f_\text{osc}}{4\pi c m_0 n},
\end{align}
where $m_0$ is the electron mass, $c$ the velocity of light, $n$ the optical index, $e$ the elementary charge, $f_\text{osc}$ the oscillator strength of the free exciton, and $E_1$ the exciton kinetic energy above which free excitons are dark.

Combining Eqs.~(\ref{eqn:equationA1})--(\ref{eqn:equationA4}), the radiative lifetime of the exciton bound to the BSF becomes $\tau_\text{r}^\text{BSF} = A T$, where $A$ is given by Eq. (\ref{eqn:slope}).

%



\begin{thebibliography}{54}%
\makeatletter
\providecommand \@ifxundefined [1]{%
 \@ifx{#1\undefined}
}%
\providecommand \@ifnum [1]{%
 \ifnum #1\expandafter \@firstoftwo
 \else \expandafter \@secondoftwo
 \fi
}%
\providecommand \@ifx [1]{%
 \ifx #1\expandafter \@firstoftwo
 \else \expandafter \@secondoftwo
 \fi
}%
\providecommand \natexlab [1]{#1}%
\providecommand \enquote  [1]{``#1''}%
\providecommand \bibnamefont  [1]{#1}%
\providecommand \bibfnamefont [1]{#1}%
\providecommand \citenamefont [1]{#1}%
\providecommand \href@noop [0]{\@secondoftwo}%
\providecommand \href [0]{\begingroup \@sanitize@url \@href}%
\providecommand \@href[1]{\@@startlink{#1}\@@href}%
\providecommand \@@href[1]{\endgroup#1\@@endlink}%
\providecommand \@sanitize@url [0]{\catcode `\\12\catcode `\$12\catcode
  `\&12\catcode `\#12\catcode `\^12\catcode `\_12\catcode `\%12\relax}%
\providecommand \@@startlink[1]{}%
\providecommand \@@endlink[0]{}%
\providecommand \url  [0]{\begingroup\@sanitize@url \@url }%
\providecommand \@url [1]{\endgroup\@href {#1}{\urlprefix }}%
\providecommand \urlprefix  [0]{URL }%
\providecommand \Eprint [0]{\href }%
\providecommand \doibase [0]{http://dx.doi.org/}%
\providecommand \selectlanguage [0]{\@gobble}%
\providecommand \bibinfo  [0]{\@secondoftwo}%
\providecommand \bibfield  [0]{\@secondoftwo}%
\providecommand \translation [1]{[#1]}%
\providecommand \BibitemOpen [0]{}%
\providecommand \bibitemStop [0]{}%
\providecommand \bibitemNoStop [0]{.\EOS\space}%
\providecommand \EOS [0]{\spacefactor3000\relax}%
\providecommand \BibitemShut  [1]{\csname bibitem#1\endcsname}%
\let\auto@bib@innerbib\@empty
\bibitem [{\citenamefont {Glas}\ \emph {et~al.}(2007)\citenamefont {Glas},
  \citenamefont {Harmand},\ and\ \citenamefont {Patriarche}}]{Glas2007}%
  \BibitemOpen
  \bibfield  {author} {\bibinfo {author} {\bibfnamefont {F.}~\bibnamefont
  {Glas}}, \bibinfo {author} {\bibfnamefont {J.-C.}\ \bibnamefont {Harmand}}, \
  and\ \bibinfo {author} {\bibfnamefont {G.}~\bibnamefont {Patriarche}},\
  }\href {\doibase 10.1103/PhysRevLett.99.146101} {\bibfield  {journal}
  {\bibinfo  {journal} {Phys. Rev. Lett.}\ }\textbf {\bibinfo {volume} {99}},\
  \bibinfo {pages} {146101} (\bibinfo {year} {2007})}\BibitemShut {NoStop}%
\bibitem [{\citenamefont {Caroff}\ \emph {et~al.}(2008)\citenamefont {Caroff},
  \citenamefont {Dick}, \citenamefont {Johansson},\ and\ \citenamefont
  {Messing}}]{Caroff2008}%
  \BibitemOpen
  \bibfield  {author} {\bibinfo {author} {\bibfnamefont {P.}~\bibnamefont
  {Caroff}}, \bibinfo {author} {\bibfnamefont {K.~A.}\ \bibnamefont {Dick}},
  \bibinfo {author} {\bibfnamefont {J.}~\bibnamefont {Johansson}}, \ and\
  \bibinfo {author} {\bibfnamefont {M.~E.}\ \bibnamefont {Messing}},\ }\href
  {\doibase 10.1038/NNANO.2008.359} {\bibfield  {journal} {\bibinfo  {journal}
  {Nature Nanotechnol.}\ }\textbf {\bibinfo {volume} {4}},\ \bibinfo {pages}
  {50} (\bibinfo {year} {2008})}\BibitemShut {NoStop}%
\bibitem [{\citenamefont {Algra}\ \emph {et~al.}(2008)\citenamefont {Algra},
  \citenamefont {Verheijen}, \citenamefont {Borgstr\"{o}m}, \citenamefont
  {Feiner}, \citenamefont {Immink}, \citenamefont {van Enckevort},
  \citenamefont {Vlieg},\ and\ \citenamefont {Bakkers}}]{Algra2008}%
  \BibitemOpen
  \bibfield  {author} {\bibinfo {author} {\bibfnamefont {R.~E.}\ \bibnamefont
  {Algra}}, \bibinfo {author} {\bibfnamefont {M.~A.}\ \bibnamefont
  {Verheijen}}, \bibinfo {author} {\bibfnamefont {M.~T.}\ \bibnamefont
  {Borgstr\"{o}m}}, \bibinfo {author} {\bibfnamefont {L.-F.}\ \bibnamefont
  {Feiner}}, \bibinfo {author} {\bibfnamefont {G.}~\bibnamefont {Immink}},
  \bibinfo {author} {\bibfnamefont {W.~J.~P.}\ \bibnamefont {van Enckevort}},
  \bibinfo {author} {\bibfnamefont {E.}~\bibnamefont {Vlieg}}, \ and\ \bibinfo
  {author} {\bibfnamefont {E.~P. A.~M.}\ \bibnamefont {Bakkers}},\ }\href
  {\doibase doi:10.1038/nature07570} {\bibfield  {journal} {\bibinfo  {journal}
  {Nature}\ }\textbf {\bibinfo {volume} {456}},\ \bibinfo {pages} {369}
  (\bibinfo {year} {2008})}\BibitemShut {NoStop}%
\bibitem [{\citenamefont {Akopian}\ \emph {et~al.}(2010)\citenamefont
  {Akopian}, \citenamefont {Patriarche}, \citenamefont {Liu}, \citenamefont
  {Harmand},\ and\ \citenamefont {Zwiller}}]{Akopian2010}%
  \BibitemOpen
  \bibfield  {author} {\bibinfo {author} {\bibfnamefont {N.}~\bibnamefont
  {Akopian}}, \bibinfo {author} {\bibfnamefont {G.}~\bibnamefont {Patriarche}},
  \bibinfo {author} {\bibfnamefont {L.}~\bibnamefont {Liu}}, \bibinfo {author}
  {\bibfnamefont {J.-C.}\ \bibnamefont {Harmand}}, \ and\ \bibinfo {author}
  {\bibfnamefont {V.}~\bibnamefont {Zwiller}},\ }\href {\doibase
  10.1021/nl903534n} {\bibfield  {journal} {\bibinfo  {journal} {Nano Lett.}\
  }\textbf {\bibinfo {volume} {10}},\ \bibinfo {pages} {1198} (\bibinfo {year}
  {2010})}\BibitemShut {NoStop}%
\bibitem [{\citenamefont {Corfdir}\ \emph {et~al.}(2013)\citenamefont
  {Corfdir}, \citenamefont {{Van Hattem}}, \citenamefont {Uccelli},
  \citenamefont {Conesa-Boj}, \citenamefont {Lefebvre}, \citenamefont
  {{Fontcuberta i Morral}},\ and\ \citenamefont {Phillips}}]{Corfdir2013}%
  \BibitemOpen
  \bibfield  {author} {\bibinfo {author} {\bibfnamefont {P.}~\bibnamefont
  {Corfdir}}, \bibinfo {author} {\bibfnamefont {B.}~\bibnamefont {{Van
  Hattem}}}, \bibinfo {author} {\bibfnamefont {E.}~\bibnamefont {Uccelli}},
  \bibinfo {author} {\bibfnamefont {S.}~\bibnamefont {Conesa-Boj}}, \bibinfo
  {author} {\bibfnamefont {P.}~\bibnamefont {Lefebvre}}, \bibinfo {author}
  {\bibfnamefont {A.}~\bibnamefont {{Fontcuberta i Morral}}}, \ and\ \bibinfo
  {author} {\bibfnamefont {R.~T.}\ \bibnamefont {Phillips}},\ }\href {\doibase
  10.1021/nl4028186} {\bibfield  {journal} {\bibinfo  {journal} {Nano Lett.}\
  }\textbf {\bibinfo {volume} {13}},\ \bibinfo {pages} {5303} (\bibinfo {year}
  {2013})}\BibitemShut {NoStop}%
\bibitem [{\citenamefont {Bolinsson}\ \emph {et~al.}(2011)\citenamefont
  {Bolinsson}, \citenamefont {Caroff}, \citenamefont {Mandl},\ and\
  \citenamefont {Dick}}]{Bolinsson2011}%
  \BibitemOpen
  \bibfield  {author} {\bibinfo {author} {\bibfnamefont {J.}~\bibnamefont
  {Bolinsson}}, \bibinfo {author} {\bibfnamefont {P.}~\bibnamefont {Caroff}},
  \bibinfo {author} {\bibfnamefont {B.}~\bibnamefont {Mandl}}, \ and\ \bibinfo
  {author} {\bibfnamefont {K.~A.}\ \bibnamefont {Dick}},\ }\href {\doibase
  10.1088/0957-4484/22/26/265606} {\bibfield  {journal} {\bibinfo  {journal}
  {Nanotechnology}\ }\textbf {\bibinfo {volume} {22}},\ \bibinfo {pages}
  {265606} (\bibinfo {year} {2011})}\BibitemShut {NoStop}%
\bibitem [{\citenamefont {Jacopin}\ \emph {et~al.}(2011)\citenamefont
  {Jacopin}, \citenamefont {Rigutti}, \citenamefont {Largeau}, \citenamefont
  {Fortuna}, \citenamefont {Furtmayr}, \citenamefont {Julien}, \citenamefont
  {Eickhoff},\ and\ \citenamefont {Tchernycheva}}]{Jacopin2011}%
  \BibitemOpen
  \bibfield  {author} {\bibinfo {author} {\bibfnamefont {G.}~\bibnamefont
  {Jacopin}}, \bibinfo {author} {\bibfnamefont {L.}~\bibnamefont {Rigutti}},
  \bibinfo {author} {\bibfnamefont {L.}~\bibnamefont {Largeau}}, \bibinfo
  {author} {\bibfnamefont {F.}~\bibnamefont {Fortuna}}, \bibinfo {author}
  {\bibfnamefont {F.}~\bibnamefont {Furtmayr}}, \bibinfo {author}
  {\bibfnamefont {F.~H.}\ \bibnamefont {Julien}}, \bibinfo {author}
  {\bibfnamefont {M.}~\bibnamefont {Eickhoff}}, \ and\ \bibinfo {author}
  {\bibfnamefont {M.}~\bibnamefont {Tchernycheva}},\ }\href
  {http://link.aip.org/link/?JAPIAU/110/064313/1} {\bibfield  {journal}
  {\bibinfo  {journal} {J. Appl. Phys.}\ }\textbf {\bibinfo {volume} {110}},\
  \bibinfo {pages} {064313} (\bibinfo {year} {2011})}\BibitemShut {NoStop}%
\bibitem [{\citenamefont {L\"ahnemann}\ \emph {et~al.}(2012)\citenamefont
  {L\"ahnemann}, \citenamefont {Brandt}, \citenamefont {Jahn}, \citenamefont
  {Pf\"uller}, \citenamefont {Roder}, \citenamefont {Dogan}, \citenamefont
  {Grosse}, \citenamefont {Belabbes}, \citenamefont {Bechstedt}, \citenamefont
  {Trampert},\ and\ \citenamefont {Geelhaar}}]{Lahnemann2012}%
  \BibitemOpen
  \bibfield  {author} {\bibinfo {author} {\bibfnamefont {J.}~\bibnamefont
  {L\"ahnemann}}, \bibinfo {author} {\bibfnamefont {O.}~\bibnamefont {Brandt}},
  \bibinfo {author} {\bibfnamefont {U.}~\bibnamefont {Jahn}}, \bibinfo {author}
  {\bibfnamefont {C.}~\bibnamefont {Pf\"uller}}, \bibinfo {author}
  {\bibfnamefont {C.}~\bibnamefont {Roder}}, \bibinfo {author} {\bibfnamefont
  {P.}~\bibnamefont {Dogan}}, \bibinfo {author} {\bibfnamefont
  {F.}~\bibnamefont {Grosse}}, \bibinfo {author} {\bibfnamefont
  {A.}~\bibnamefont {Belabbes}}, \bibinfo {author} {\bibfnamefont
  {F.}~\bibnamefont {Bechstedt}}, \bibinfo {author} {\bibfnamefont
  {A.}~\bibnamefont {Trampert}}, \ and\ \bibinfo {author} {\bibfnamefont
  {L.}~\bibnamefont {Geelhaar}},\ }\href {\doibase 10.1103/PhysRevB.86.081302}
  {\bibfield  {journal} {\bibinfo  {journal} {Phys. Rev. B}\ }\textbf {\bibinfo
  {volume} {86}},\ \bibinfo {pages} {081302} (\bibinfo {year}
  {2012})}\BibitemShut {NoStop}%
\bibitem [{\citenamefont {Graham}\ \emph {et~al.}(2013)\citenamefont {Graham},
  \citenamefont {Corfdir}, \citenamefont {Heiss}, \citenamefont {Conesa-Boj},
  \citenamefont {Uccelli}, \citenamefont {Fontcuberta~i Morral},\ and\
  \citenamefont {Phillips}}]{Graham2013}%
  \BibitemOpen
  \bibfield  {author} {\bibinfo {author} {\bibfnamefont {A.~M.}\ \bibnamefont
  {Graham}}, \bibinfo {author} {\bibfnamefont {P.}~\bibnamefont {Corfdir}},
  \bibinfo {author} {\bibfnamefont {M.}~\bibnamefont {Heiss}}, \bibinfo
  {author} {\bibfnamefont {S.}~\bibnamefont {Conesa-Boj}}, \bibinfo {author}
  {\bibfnamefont {E.}~\bibnamefont {Uccelli}}, \bibinfo {author} {\bibfnamefont
  {A.}~\bibnamefont {Fontcuberta~i Morral}}, \ and\ \bibinfo {author}
  {\bibfnamefont {R.~T.}\ \bibnamefont {Phillips}},\ }\href {\doibase
  10.1103/PhysRevB.87.125304} {\bibfield  {journal} {\bibinfo  {journal} {Phys.
  Rev. B}\ }\textbf {\bibinfo {volume} {87}},\ \bibinfo {pages} {125304}
  (\bibinfo {year} {2013})}\BibitemShut {NoStop}%
\bibitem [{\citenamefont {Castelletto}\ \emph {et~al.}(2014)\citenamefont
  {Castelletto}, \citenamefont {Bodrog}, \citenamefont {Magyar}, \citenamefont
  {Gentle}, \citenamefont {Gali},\ and\ \citenamefont
  {Aharonovich}}]{Castelletto2014}%
  \BibitemOpen
  \bibfield  {author} {\bibinfo {author} {\bibfnamefont {S.}~\bibnamefont
  {Castelletto}}, \bibinfo {author} {\bibfnamefont {Z.}~\bibnamefont {Bodrog}},
  \bibinfo {author} {\bibfnamefont {A.~P.}\ \bibnamefont {Magyar}}, \bibinfo
  {author} {\bibfnamefont {A.}~\bibnamefont {Gentle}}, \bibinfo {author}
  {\bibfnamefont {A.}~\bibnamefont {Gali}}, \ and\ \bibinfo {author}
  {\bibfnamefont {I.}~\bibnamefont {Aharonovich}},\ }\href {\doibase
  10.1039/c4nr02307b} {\bibfield  {journal} {\bibinfo  {journal} {Nanoscale}\ }
  (\bibinfo {year} {2014}),\ 10.1039/c4nr02307b}\BibitemShut {NoStop}%
\bibitem [{\citenamefont {Kouno}\ \emph {et~al.}(2014)\citenamefont {Kouno},
  \citenamefont {Sakai}, \citenamefont {Kishino},\ and\ \citenamefont
  {Hara}}]{Kouno2014}%
  \BibitemOpen
  \bibfield  {author} {\bibinfo {author} {\bibfnamefont {T.}~\bibnamefont
  {Kouno}}, \bibinfo {author} {\bibfnamefont {M.}~\bibnamefont {Sakai}},
  \bibinfo {author} {\bibfnamefont {K.}~\bibnamefont {Kishino}}, \ and\
  \bibinfo {author} {\bibfnamefont {K.}~\bibnamefont {Hara}},\ }\href {\doibase
  10.7567/JJAP.53.068001} {\bibfield  {journal} {\bibinfo  {journal} {Jpn. J.
  Appl. Phys.}\ }\textbf {\bibinfo {volume} {53}},\ \bibinfo {pages} {068001}
  (\bibinfo {year} {2014})}\BibitemShut {NoStop}%
\bibitem [{\citenamefont {Vu}\ \emph {et~al.}(2013)\citenamefont {Vu},
  \citenamefont {Zehender}, \citenamefont {Verheijen}, \citenamefont
  {Plissard}, \citenamefont {Immink}, \citenamefont {Haverkort},\ and\
  \citenamefont {Bakkers}}]{Vu2013}%
  \BibitemOpen
  \bibfield  {author} {\bibinfo {author} {\bibfnamefont {T.~T.~T.}\
  \bibnamefont {Vu}}, \bibinfo {author} {\bibfnamefont {T.}~\bibnamefont
  {Zehender}}, \bibinfo {author} {\bibfnamefont {M.~A.}\ \bibnamefont
  {Verheijen}}, \bibinfo {author} {\bibfnamefont {S.~R.}\ \bibnamefont
  {Plissard}}, \bibinfo {author} {\bibfnamefont {G.~W.~G.}\ \bibnamefont
  {Immink}}, \bibinfo {author} {\bibfnamefont {J.~E.~M.}\ \bibnamefont
  {Haverkort}}, \ and\ \bibinfo {author} {\bibfnamefont {E.~P. A.~M.}\
  \bibnamefont {Bakkers}},\ }\href {\doibase 10.1088/0957-4484/24/11/115705}
  {\bibfield  {journal} {\bibinfo  {journal} {Nanotechnol.}\ }\textbf {\bibinfo
  {volume} {24}},\ \bibinfo {pages} {115705} (\bibinfo {year}
  {2013})}\BibitemShut {NoStop}%
\bibitem [{\citenamefont {Ahtapodov}\ \emph {et~al.}(2012)\citenamefont
  {Ahtapodov}, \citenamefont {Todorovic}, \citenamefont {Olk}, \citenamefont
  {Mj\r{a}land}, \citenamefont {Sl\r{a}ttnes}, \citenamefont {Dheeraj},
  \citenamefont {Helvoort}, \citenamefont {Fimland},\ and\ \citenamefont
  {Weman}}]{Ahtapodov2012}%
  \BibitemOpen
  \bibfield  {author} {\bibinfo {author} {\bibfnamefont {L.}~\bibnamefont
  {Ahtapodov}}, \bibinfo {author} {\bibfnamefont {J.}~\bibnamefont
  {Todorovic}}, \bibinfo {author} {\bibfnamefont {P.}~\bibnamefont {Olk}},
  \bibinfo {author} {\bibfnamefont {T.}~\bibnamefont {Mj\r{a}land}}, \bibinfo
  {author} {\bibfnamefont {P.}~\bibnamefont {Sl\r{a}ttnes}}, \bibinfo {author}
  {\bibfnamefont {D.~L.}\ \bibnamefont {Dheeraj}}, \bibinfo {author}
  {\bibfnamefont {A.~T.~J.}\ \bibnamefont {Helvoort}}, \bibinfo {author}
  {\bibfnamefont {B.-O.}\ \bibnamefont {Fimland}}, \ and\ \bibinfo {author}
  {\bibfnamefont {H.}~\bibnamefont {Weman}},\ }\href {\doibase
  10.1021/nl3025714} {\bibfield  {journal} {\bibinfo  {journal} {Nano Lett.}\
  }\textbf {\bibinfo {volume} {12}},\ \bibinfo {pages} {6090} (\bibinfo {year}
  {2012})}\BibitemShut {NoStop}%
\bibitem [{\citenamefont {Albrecht}\ \emph {et~al.}(1997)\citenamefont
  {Albrecht}, \citenamefont {Christiansen}, \citenamefont {Salviati},
  \citenamefont {Zanotti-Fregonara}, \citenamefont {Rebane}, \citenamefont
  {Shreter}, \citenamefont {Mayer}, \citenamefont {Pelzmann}, \citenamefont
  {Kamp}, \citenamefont {Ebeling}, \citenamefont {Bremser}, \citenamefont
  {Davis},\ and\ \citenamefont {Strunk}}]{Albrecht1997}%
  \BibitemOpen
  \bibfield  {author} {\bibinfo {author} {\bibfnamefont {M.}~\bibnamefont
  {Albrecht}}, \bibinfo {author} {\bibfnamefont {S.}~\bibnamefont
  {Christiansen}}, \bibinfo {author} {\bibfnamefont {G.}~\bibnamefont
  {Salviati}}, \bibinfo {author} {\bibfnamefont {C.}~\bibnamefont
  {Zanotti-Fregonara}}, \bibinfo {author} {\bibfnamefont {Y.~T.}\ \bibnamefont
  {Rebane}}, \bibinfo {author} {\bibfnamefont {Y.~G.}\ \bibnamefont {Shreter}},
  \bibinfo {author} {\bibfnamefont {M.}~\bibnamefont {Mayer}}, \bibinfo
  {author} {\bibfnamefont {A.}~\bibnamefont {Pelzmann}}, \bibinfo {author}
  {\bibfnamefont {M.}~\bibnamefont {Kamp}}, \bibinfo {author} {\bibfnamefont
  {K.~J.}\ \bibnamefont {Ebeling}}, \bibinfo {author} {\bibfnamefont {M.~D.}\
  \bibnamefont {Bremser}}, \bibinfo {author} {\bibfnamefont {R.~F.}\
  \bibnamefont {Davis}}, \ and\ \bibinfo {author} {\bibfnamefont {H.~P.}\
  \bibnamefont {Strunk}},\ }\href {\doibase 10.1557/PROC-468-293} {\bibfield
  {journal} {\bibinfo  {journal} {MRS Symp. Proc.}\ }\textbf {\bibinfo {volume}
  {468}},\ \bibinfo {pages} {293} (\bibinfo {year} {1997})}\BibitemShut
  {NoStop}%
\bibitem [{\citenamefont {Stampfl}\ and\ \citenamefont {Van~de
  Walle}(1998)}]{Stampfl1998}%
  \BibitemOpen
  \bibfield  {author} {\bibinfo {author} {\bibfnamefont {C.}~\bibnamefont
  {Stampfl}}\ and\ \bibinfo {author} {\bibfnamefont {C.~G.}\ \bibnamefont
  {Van~de Walle}},\ }\href {\doibase 10.1103/PhysRevB.57.R15052} {\bibfield
  {journal} {\bibinfo  {journal} {Phys. Rev. B}\ }\textbf {\bibinfo {volume}
  {57}},\ \bibinfo {pages} {R15052} (\bibinfo {year} {1998})}\BibitemShut
  {NoStop}%
\bibitem [{\citenamefont {Korona}\ \emph {et~al.}(2014)\citenamefont {Korona},
  \citenamefont {Reszka}, \citenamefont {Sobanska}, \citenamefont {Perkowska},
  \citenamefont {Wysmolek}, \citenamefont {Klosek},\ and\ \citenamefont
  {Zytkiewicz}}]{Korona2014}%
  \BibitemOpen
  \bibfield  {author} {\bibinfo {author} {\bibfnamefont {K.}~\bibnamefont
  {Korona}}, \bibinfo {author} {\bibfnamefont {A.}~\bibnamefont {Reszka}},
  \bibinfo {author} {\bibfnamefont {M.}~\bibnamefont {Sobanska}}, \bibinfo
  {author} {\bibfnamefont {P.}~\bibnamefont {Perkowska}}, \bibinfo {author}
  {\bibfnamefont {A.}~\bibnamefont {Wysmolek}}, \bibinfo {author}
  {\bibfnamefont {K.}~\bibnamefont {Klosek}}, \ and\ \bibinfo {author}
  {\bibfnamefont {Z.}~\bibnamefont {Zytkiewicz}},\ }\href {\doibase
  10.1016/j.jlumin.2014.06.061} {\bibfield  {journal} {\bibinfo  {journal} {J.
  Lumin.}\ }\textbf {\bibinfo {volume} {155}},\ \bibinfo {pages} {293}
  (\bibinfo {year} {2014})}\BibitemShut {NoStop}%
\bibitem [{\citenamefont {Fern\'{a}ndez-Garrido}\ \emph
  {et~al.}(2009)\citenamefont {Fern\'{a}ndez-Garrido}, \citenamefont {Grandal},
  \citenamefont {Calleja}, \citenamefont {S\'{a}nchez-Garc\'{\i}a},\ and\
  \citenamefont {Lopez-Romero}}]{Garrido2009}%
  \BibitemOpen
  \bibfield  {author} {\bibinfo {author} {\bibfnamefont {S.}~\bibnamefont
  {Fern\'{a}ndez-Garrido}}, \bibinfo {author} {\bibfnamefont {J.}~\bibnamefont
  {Grandal}}, \bibinfo {author} {\bibfnamefont {E.}~\bibnamefont {Calleja}},
  \bibinfo {author} {\bibfnamefont {M.~A.}\ \bibnamefont
  {S\'{a}nchez-Garc\'{\i}a}}, \ and\ \bibinfo {author} {\bibfnamefont
  {D.}~\bibnamefont {Lopez-Romero}},\ }\href {\doibase 10.1063/1.3267151}
  {\bibfield  {journal} {\bibinfo  {journal} {J. Appl. Phys.}\ }\textbf
  {\bibinfo {volume} {106}},\ \bibinfo {pages} {126102} (\bibinfo {year}
  {2009})}\BibitemShut {NoStop}%
\bibitem [{\citenamefont {Geelhaar}\ \emph {et~al.}(2011)\citenamefont
  {Geelhaar}, \citenamefont {Ch\`{e}ze}, \citenamefont {Jenichen},
  \citenamefont {Brandt}, \citenamefont {Pf\"{u}ller}, \citenamefont
  {M\"{u}nch}, \citenamefont {Rothemund}, \citenamefont {Reitzenstein},
  \citenamefont {Forchel}, \citenamefont {Kehagias}, \citenamefont {Komninou},
  \citenamefont {Dimitrakopulos}, \citenamefont {Karakostas}, \citenamefont
  {Lari}, \citenamefont {Chalker}, \citenamefont {Gass},\ and\ \citenamefont
  {Riechert}}]{Geelhaar2011}%
  \BibitemOpen
  \bibfield  {author} {\bibinfo {author} {\bibfnamefont {L.}~\bibnamefont
  {Geelhaar}}, \bibinfo {author} {\bibfnamefont {C.}~\bibnamefont {Ch\`{e}ze}},
  \bibinfo {author} {\bibfnamefont {B.}~\bibnamefont {Jenichen}}, \bibinfo
  {author} {\bibfnamefont {O.}~\bibnamefont {Brandt}}, \bibinfo {author}
  {\bibfnamefont {C.}~\bibnamefont {Pf\"{u}ller}}, \bibinfo {author}
  {\bibfnamefont {S.}~\bibnamefont {M\"{u}nch}}, \bibinfo {author}
  {\bibfnamefont {R.}~\bibnamefont {Rothemund}}, \bibinfo {author}
  {\bibfnamefont {S.}~\bibnamefont {Reitzenstein}}, \bibinfo {author}
  {\bibfnamefont {A.}~\bibnamefont {Forchel}}, \bibinfo {author} {\bibfnamefont
  {T.}~\bibnamefont {Kehagias}}, \bibinfo {author} {\bibfnamefont
  {P.}~\bibnamefont {Komninou}}, \bibinfo {author} {\bibfnamefont {G.~P.}\
  \bibnamefont {Dimitrakopulos}}, \bibinfo {author} {\bibfnamefont
  {T.}~\bibnamefont {Karakostas}}, \bibinfo {author} {\bibfnamefont
  {L.}~\bibnamefont {Lari}}, \bibinfo {author} {\bibfnamefont {P.~R.}\
  \bibnamefont {Chalker}}, \bibinfo {author} {\bibfnamefont {M.~H.}\
  \bibnamefont {Gass}}, \ and\ \bibinfo {author} {\bibfnamefont
  {H.}~\bibnamefont {Riechert}},\ }\href {\doibase 10.1109/JSTQE.2010.2098396}
  {\bibfield  {journal} {\bibinfo  {journal} {IEEE J. Sel. Top. Quantum
  Electron.}\ }\textbf {\bibinfo {volume} {17}},\ \bibinfo {pages} {878}
  (\bibinfo {year} {2011})}\BibitemShut {NoStop}%
\bibitem [{\citenamefont {Consonni}(2013)}]{Consonni2013}%
  \BibitemOpen
  \bibfield  {author} {\bibinfo {author} {\bibfnamefont {V.}~\bibnamefont
  {Consonni}},\ }\href {\doibase 10.1002/pssr.201307237} {\bibfield  {journal}
  {\bibinfo  {journal} {Physica Status Solidi RRL}\ }\textbf {\bibinfo {volume}
  {7}},\ \bibinfo {pages} {699} (\bibinfo {year} {2013})}\BibitemShut {NoStop}%
\bibitem [{\citenamefont {Heying}\ \emph {et~al.}(2000)\citenamefont {Heying},
  \citenamefont {Averbeck}, \citenamefont {Chen}, \citenamefont {Haus},
  \citenamefont {Riechert},\ and\ \citenamefont {Speck}}]{Heying2000}%
  \BibitemOpen
  \bibfield  {author} {\bibinfo {author} {\bibfnamefont {B.}~\bibnamefont
  {Heying}}, \bibinfo {author} {\bibfnamefont {R.}~\bibnamefont {Averbeck}},
  \bibinfo {author} {\bibfnamefont {L.~F.}\ \bibnamefont {Chen}}, \bibinfo
  {author} {\bibfnamefont {E.}~\bibnamefont {Haus}}, \bibinfo {author}
  {\bibfnamefont {H.}~\bibnamefont {Riechert}}, \ and\ \bibinfo {author}
  {\bibfnamefont {J.~S.}\ \bibnamefont {Speck}},\ }\href {\doibase
  10.1063/1.1305830} {\bibfield  {journal} {\bibinfo  {journal} {J. Appl.
  Phys.}\ }\textbf {\bibinfo {volume} {88}},\ \bibinfo {pages} {1855} (\bibinfo
  {year} {2000})}\BibitemShut {NoStop}%
\bibitem [{\citenamefont {Corfdir}\ \emph {et~al.}(2014)\citenamefont
  {Corfdir}, \citenamefont {Zettler}, \citenamefont {Hauswald}, \citenamefont
  {Fern\'{a}ndez-Garrido}, \citenamefont {Brandt},\ and\ \citenamefont
  {Lefebvre}}]{Corfdir2014}%
  \BibitemOpen
  \bibfield  {author} {\bibinfo {author} {\bibfnamefont {P.}~\bibnamefont
  {Corfdir}}, \bibinfo {author} {\bibfnamefont {J.~K.}\ \bibnamefont
  {Zettler}}, \bibinfo {author} {\bibfnamefont {C.}~\bibnamefont {Hauswald}},
  \bibinfo {author} {\bibfnamefont {S.}~\bibnamefont {Fern\'{a}ndez-Garrido}},
  \bibinfo {author} {\bibfnamefont {O.}~\bibnamefont {Brandt}}, \ and\ \bibinfo
  {author} {\bibfnamefont {P.}~\bibnamefont {Lefebvre}},\ }\href@noop {}
  {\bibfield  {journal} {\bibinfo  {journal} {submitted to Phys. Rev. B}\ }
  (\bibinfo {year} {2014})},\ \Eprint {http://arxiv.org/abs/1407.4279}
  {arXiv:1407.4279} \BibitemShut {NoStop}%
\bibitem [{\citenamefont {Corfdir}\ \emph
  {et~al.}(2009{\natexlab{a}})\citenamefont {Corfdir}, \citenamefont
  {Lefebvre}, \citenamefont {Risti\'{c}}, \citenamefont {Valvin}, \citenamefont
  {Calleja}, \citenamefont {Trampert}, \citenamefont {Gani\`{e}re},\ and\
  \citenamefont {Deveaud-Pl\'{e}dran}}]{Corfdir2009c}%
  \BibitemOpen
  \bibfield  {author} {\bibinfo {author} {\bibfnamefont {P.}~\bibnamefont
  {Corfdir}}, \bibinfo {author} {\bibfnamefont {P.}~\bibnamefont {Lefebvre}},
  \bibinfo {author} {\bibfnamefont {J.}~\bibnamefont {Risti\'{c}}}, \bibinfo
  {author} {\bibfnamefont {P.}~\bibnamefont {Valvin}}, \bibinfo {author}
  {\bibfnamefont {E.}~\bibnamefont {Calleja}}, \bibinfo {author} {\bibfnamefont
  {A.}~\bibnamefont {Trampert}}, \bibinfo {author} {\bibfnamefont {J.-D.}\
  \bibnamefont {Gani\`{e}re}}, \ and\ \bibinfo {author} {\bibfnamefont
  {B.}~\bibnamefont {Deveaud-Pl\'{e}dran}},\ }\href {\doibase
  10.1063/1.3062742} {\bibfield  {journal} {\bibinfo  {journal} {J. Appl.
  Phys.}\ }\textbf {\bibinfo {volume} {105}},\ \bibinfo {pages} {013113}
  (\bibinfo {year} {2009}{\natexlab{a}})}\BibitemShut {NoStop}%
\bibitem [{\citenamefont {Calleja}\ \emph {et~al.}(2000)\citenamefont
  {Calleja}, \citenamefont {S\'{a}nchez-Garc\'{\i}a}, \citenamefont
  {S\'{a}nchez}, \citenamefont {Calle}, \citenamefont {Naranjo}, \citenamefont
  {Mu\~{n}oz}, \citenamefont {Jahn},\ and\ \citenamefont
  {Ploog}}]{Calleja2000}%
  \BibitemOpen
  \bibfield  {author} {\bibinfo {author} {\bibfnamefont {E.}~\bibnamefont
  {Calleja}}, \bibinfo {author} {\bibfnamefont {M.}~\bibnamefont
  {S\'{a}nchez-Garc\'{\i}a}}, \bibinfo {author} {\bibfnamefont
  {F.}~\bibnamefont {S\'{a}nchez}}, \bibinfo {author} {\bibfnamefont
  {F.}~\bibnamefont {Calle}}, \bibinfo {author} {\bibfnamefont
  {F.}~\bibnamefont {Naranjo}}, \bibinfo {author} {\bibfnamefont
  {E.}~\bibnamefont {Mu\~{n}oz}}, \bibinfo {author} {\bibfnamefont
  {U.}~\bibnamefont {Jahn}}, \ and\ \bibinfo {author} {\bibfnamefont {K.~H.}\
  \bibnamefont {Ploog}},\ }\href {\doibase 10.1103/PhysRevB.62.16826}
  {\bibfield  {journal} {\bibinfo  {journal} {Phys. Rev. B}\ }\textbf {\bibinfo
  {volume} {62}},\ \bibinfo {pages} {16826} (\bibinfo {year}
  {2000})}\BibitemShut {NoStop}%
\bibitem [{\citenamefont {Lefebvre}\ \emph {et~al.}(2011)\citenamefont
  {Lefebvre}, \citenamefont {Fern\'{a}ndez-Garrido}, \citenamefont {Grandal},
  \citenamefont {Risti\'{c}}, \citenamefont {S\'{a}nchez-Garc\'{\i}a},\ and\
  \citenamefont {Calleja}}]{Lefebvre2011}%
  \BibitemOpen
  \bibfield  {author} {\bibinfo {author} {\bibfnamefont {P.}~\bibnamefont
  {Lefebvre}}, \bibinfo {author} {\bibfnamefont {S.}~\bibnamefont
  {Fern\'{a}ndez-Garrido}}, \bibinfo {author} {\bibfnamefont {J.}~\bibnamefont
  {Grandal}}, \bibinfo {author} {\bibfnamefont {J.}~\bibnamefont {Risti\'{c}}},
  \bibinfo {author} {\bibfnamefont {M.-A.}\ \bibnamefont
  {S\'{a}nchez-Garc\'{\i}a}}, \ and\ \bibinfo {author} {\bibfnamefont
  {E.}~\bibnamefont {Calleja}},\ }\href {\doibase 10.1063/1.3556643} {\bibfield
   {journal} {\bibinfo  {journal} {App. Phys. Lett.}\ }\textbf {\bibinfo
  {volume} {98}},\ \bibinfo {pages} {083104} (\bibinfo {year}
  {2011})}\BibitemShut {NoStop}%
\bibitem [{\citenamefont {Paskov}\ \emph {et~al.}(2005)\citenamefont {Paskov},
  \citenamefont {Schifano}, \citenamefont {Monemar}, \citenamefont {Paskova},
  \citenamefont {Figge},\ and\ \citenamefont {Hommel}}]{Paskov2005}%
  \BibitemOpen
  \bibfield  {author} {\bibinfo {author} {\bibfnamefont {P.~P.}\ \bibnamefont
  {Paskov}}, \bibinfo {author} {\bibfnamefont {R.}~\bibnamefont {Schifano}},
  \bibinfo {author} {\bibfnamefont {B.}~\bibnamefont {Monemar}}, \bibinfo
  {author} {\bibfnamefont {T.}~\bibnamefont {Paskova}}, \bibinfo {author}
  {\bibfnamefont {S.}~\bibnamefont {Figge}}, \ and\ \bibinfo {author}
  {\bibfnamefont {D.}~\bibnamefont {Hommel}},\ }\href {\doibase
  http://dx.doi.org/10.1063/1.2128496} {\bibfield  {journal} {\bibinfo
  {journal} {J. Appl. Phys.}\ }\textbf {\bibinfo {volume} {98}},\ \bibinfo
  {eid} {093519} (\bibinfo {year} {2005})}\BibitemShut {NoStop}%
\bibitem [{\citenamefont {G\"uhne}\ \emph {et~al.}(2008)\citenamefont
  {G\"uhne}, \citenamefont {Bougrioua}, \citenamefont {La\"ugt}, \citenamefont
  {Nemoz}, \citenamefont {Venn\'egu\`es}, \citenamefont {Vinter},\ and\
  \citenamefont {Leroux}}]{Guhne2008}%
  \BibitemOpen
  \bibfield  {author} {\bibinfo {author} {\bibfnamefont {T.}~\bibnamefont
  {G\"uhne}}, \bibinfo {author} {\bibfnamefont {Z.}~\bibnamefont {Bougrioua}},
  \bibinfo {author} {\bibfnamefont {S.}~\bibnamefont {La\"ugt}}, \bibinfo
  {author} {\bibfnamefont {M.}~\bibnamefont {Nemoz}}, \bibinfo {author}
  {\bibfnamefont {P.}~\bibnamefont {Venn\'egu\`es}}, \bibinfo {author}
  {\bibfnamefont {B.}~\bibnamefont {Vinter}}, \ and\ \bibinfo {author}
  {\bibfnamefont {M.}~\bibnamefont {Leroux}},\ }\href {\doibase
  10.1103/PhysRevB.77.075308} {\bibfield  {journal} {\bibinfo  {journal} {Phys.
  Rev. B}\ }\textbf {\bibinfo {volume} {77}},\ \bibinfo {pages} {075308}
  (\bibinfo {year} {2008})}\BibitemShut {NoStop}%
\bibitem [{\citenamefont {Corfdir}\ \emph
  {et~al.}(2009{\natexlab{b}})\citenamefont {Corfdir}, \citenamefont
  {Lefebvre}, \citenamefont {Levrat}, \citenamefont {Dussaigne}, \citenamefont
  {Gani\`{e}re}, \citenamefont {Martin}, \citenamefont {Risti\'{c}},
  \citenamefont {Zhu}, \citenamefont {Grandjean},\ and\ \citenamefont
  {Deveaud-Pl\'edran}}]{Corfdir2009a}%
  \BibitemOpen
  \bibfield  {author} {\bibinfo {author} {\bibfnamefont {P.}~\bibnamefont
  {Corfdir}}, \bibinfo {author} {\bibfnamefont {P.}~\bibnamefont {Lefebvre}},
  \bibinfo {author} {\bibfnamefont {J.}~\bibnamefont {Levrat}}, \bibinfo
  {author} {\bibfnamefont {A.}~\bibnamefont {Dussaigne}}, \bibinfo {author}
  {\bibfnamefont {J.-D.}\ \bibnamefont {Gani\`{e}re}}, \bibinfo {author}
  {\bibfnamefont {D.}~\bibnamefont {Martin}}, \bibinfo {author} {\bibfnamefont
  {J.}~\bibnamefont {Risti\'{c}}}, \bibinfo {author} {\bibfnamefont
  {T.}~\bibnamefont {Zhu}}, \bibinfo {author} {\bibfnamefont {N.}~\bibnamefont
  {Grandjean}}, \ and\ \bibinfo {author} {\bibfnamefont {B.}~\bibnamefont
  {Deveaud-Pl\'edran}},\ }\href {\doibase http://dx.doi.org/10.1063/1.3075596}
  {\bibfield  {journal} {\bibinfo  {journal} {J. Appl. Phys.}\ }\textbf
  {\bibinfo {volume} {105}},\ \bibinfo {pages} {043102} (\bibinfo {year}
  {2009}{\natexlab{b}})}\BibitemShut {NoStop}%
\bibitem [{\citenamefont {Corfdir}\ and\ \citenamefont
  {Lefebvre}(2012)}]{Corfdir2012}%
  \BibitemOpen
  \bibfield  {author} {\bibinfo {author} {\bibfnamefont {P.}~\bibnamefont
  {Corfdir}}\ and\ \bibinfo {author} {\bibfnamefont {P.}~\bibnamefont
  {Lefebvre}},\ }\href {\doibase http://dx.doi.org/10.1063/1.4749789}
  {\bibfield  {journal} {\bibinfo  {journal} {J. Appl. Phys.}\ }\textbf
  {\bibinfo {volume} {112}},\ \bibinfo {eid} {053512} (\bibinfo {year}
  {2012})}\BibitemShut {NoStop}%
\bibitem [{\citenamefont {Chisholm}\ and\ \citenamefont
  {Bristowe}(2000)}]{Chisholm2000}%
  \BibitemOpen
  \bibfield  {author} {\bibinfo {author} {\bibfnamefont {J.~A.}\ \bibnamefont
  {Chisholm}}\ and\ \bibinfo {author} {\bibfnamefont {P.~D.}\ \bibnamefont
  {Bristowe}},\ }\href {\doibase 10.1063/1.127035} {\bibfield  {journal}
  {\bibinfo  {journal} {Applied Physics Letters}\ }\textbf {\bibinfo {volume}
  {77}},\ \bibinfo {pages} {534} (\bibinfo {year} {2000})}\BibitemShut
  {NoStop}%
\bibitem [{\citenamefont {Corfdir}\ \emph
  {et~al.}(2009{\natexlab{c}})\citenamefont {Corfdir}, \citenamefont
  {Risti\'{c}}, \citenamefont {Lefebvre}, \citenamefont {Zhu}, \citenamefont
  {Martin}, \citenamefont {Dussaigne}, \citenamefont {Gani\`{e}re},
  \citenamefont {Grandjean},\ and\ \citenamefont
  {Deveaud-Pl\'edran}}]{Corfdir2009b}%
  \BibitemOpen
  \bibfield  {author} {\bibinfo {author} {\bibfnamefont {P.}~\bibnamefont
  {Corfdir}}, \bibinfo {author} {\bibfnamefont {J.}~\bibnamefont {Risti\'{c}}},
  \bibinfo {author} {\bibfnamefont {P.}~\bibnamefont {Lefebvre}}, \bibinfo
  {author} {\bibfnamefont {T.}~\bibnamefont {Zhu}}, \bibinfo {author}
  {\bibfnamefont {D.}~\bibnamefont {Martin}}, \bibinfo {author} {\bibfnamefont
  {A.}~\bibnamefont {Dussaigne}}, \bibinfo {author} {\bibfnamefont {J.~D.}\
  \bibnamefont {Gani\`{e}re}}, \bibinfo {author} {\bibfnamefont
  {N.}~\bibnamefont {Grandjean}}, \ and\ \bibinfo {author} {\bibfnamefont
  {B.}~\bibnamefont {Deveaud-Pl\'edran}},\ }\href {\doibase
  http://dx.doi.org/10.1063/1.3142396} {\bibfield  {journal} {\bibinfo
  {journal} {Appl. Phys. Lett.}\ }\textbf {\bibinfo {volume} {94}},\ \bibinfo
  {pages} {201115} (\bibinfo {year} {2009}{\natexlab{c}})}\BibitemShut
  {NoStop}%
\bibitem [{\citenamefont {Corfdir}\ \emph {et~al.}(2011)\citenamefont
  {Corfdir}, \citenamefont {Levrat}, \citenamefont {Dussaigne}, \citenamefont
  {Lefebvre}, \citenamefont {Teisseyre}, \citenamefont {Grzegory},
  \citenamefont {Suski}, \citenamefont {Gani\`{e}re}, \citenamefont
  {Grandjean},\ and\ \citenamefont {Deveaud-Pl\'{e}dran}}]{Corfdir2011}%
  \BibitemOpen
  \bibfield  {author} {\bibinfo {author} {\bibfnamefont {P.}~\bibnamefont
  {Corfdir}}, \bibinfo {author} {\bibfnamefont {J.}~\bibnamefont {Levrat}},
  \bibinfo {author} {\bibfnamefont {A.}~\bibnamefont {Dussaigne}}, \bibinfo
  {author} {\bibfnamefont {P.}~\bibnamefont {Lefebvre}}, \bibinfo {author}
  {\bibfnamefont {H.}~\bibnamefont {Teisseyre}}, \bibinfo {author}
  {\bibfnamefont {I.}~\bibnamefont {Grzegory}}, \bibinfo {author}
  {\bibfnamefont {T.}~\bibnamefont {Suski}}, \bibinfo {author} {\bibfnamefont
  {J.-D.}\ \bibnamefont {Gani\`{e}re}}, \bibinfo {author} {\bibfnamefont
  {N.}~\bibnamefont {Grandjean}}, \ and\ \bibinfo {author} {\bibfnamefont
  {B.}~\bibnamefont {Deveaud-Pl\'{e}dran}},\ }\href {\doibase
  10.1103/PhysRevB.83.245326} {\bibfield  {journal} {\bibinfo  {journal} {Phys.
  Rev. B}\ }\textbf {\bibinfo {volume} {83}},\ \bibinfo {pages} {245326}
  (\bibinfo {year} {2011})}\BibitemShut {NoStop}%
\bibitem [{\citenamefont {Nogues}\ \emph {et~al.}(2014)\citenamefont {Nogues},
  \citenamefont {Auzelle}, \citenamefont {{Den Hertog}}, \citenamefont
  {Gayral},\ and\ \citenamefont {Daudin}}]{Nogues2014}%
  \BibitemOpen
  \bibfield  {author} {\bibinfo {author} {\bibfnamefont {G.}~\bibnamefont
  {Nogues}}, \bibinfo {author} {\bibfnamefont {T.}~\bibnamefont {Auzelle}},
  \bibinfo {author} {\bibfnamefont {M.}~\bibnamefont {{Den Hertog}}}, \bibinfo
  {author} {\bibfnamefont {B.}~\bibnamefont {Gayral}}, \ and\ \bibinfo {author}
  {\bibfnamefont {B.}~\bibnamefont {Daudin}},\ }\href {\doibase
  10.1063/1.4868131} {\bibfield  {journal} {\bibinfo  {journal} {Appl. Phys.
  Lett.}\ }\textbf {\bibinfo {volume} {104}},\ \bibinfo {pages} {102102}
  (\bibinfo {year} {2014})}\BibitemShut {NoStop}%
\bibitem [{\citenamefont {P\"assler}(2001)}]{Paessler2001}%
  \BibitemOpen
  \bibfield  {author} {\bibinfo {author} {\bibfnamefont {R.}~\bibnamefont
  {P\"assler}},\ }\href {\doibase http://dx.doi.org/10.1063/1.1402147}
  {\bibfield  {journal} {\bibinfo  {journal} {J. Appl. Phys.}\ }\textbf
  {\bibinfo {volume} {90}},\ \bibinfo {pages} {3956} (\bibinfo {year}
  {2001})}\BibitemShut {NoStop}%
\bibitem [{\citenamefont {Hauswald}\ \emph
  {et~al.}(2014{\natexlab{a}})\citenamefont {Hauswald}, \citenamefont
  {Corfdir}, \citenamefont {Zettler}, \citenamefont {Kaganer}, \citenamefont
  {Sabelfeld}, \citenamefont {Fern\'{a}ndez-Garrido}, \citenamefont
  {Flissikowski}, \citenamefont {Consonni}, \citenamefont {Gotschke},
  \citenamefont {Grahn}, \citenamefont {Geelhaar},\ and\ \citenamefont
  {Brandt}}]{Hauswald2014b}%
  \BibitemOpen
  \bibfield  {author} {\bibinfo {author} {\bibfnamefont {C.}~\bibnamefont
  {Hauswald}}, \bibinfo {author} {\bibfnamefont {P.}~\bibnamefont {Corfdir}},
  \bibinfo {author} {\bibfnamefont {J.~K.}\ \bibnamefont {Zettler}}, \bibinfo
  {author} {\bibfnamefont {V.}~\bibnamefont {Kaganer}}, \bibinfo {author}
  {\bibfnamefont {K.~K.}\ \bibnamefont {Sabelfeld}}, \bibinfo {author}
  {\bibfnamefont {S.}~\bibnamefont {Fern\'{a}ndez-Garrido}}, \bibinfo {author}
  {\bibfnamefont {T.}~\bibnamefont {Flissikowski}}, \bibinfo {author}
  {\bibfnamefont {V.}~\bibnamefont {Consonni}}, \bibinfo {author}
  {\bibfnamefont {T.}~\bibnamefont {Gotschke}}, \bibinfo {author}
  {\bibfnamefont {H.~T.}\ \bibnamefont {Grahn}}, \bibinfo {author}
  {\bibfnamefont {L.}~\bibnamefont {Geelhaar}}, \ and\ \bibinfo {author}
  {\bibfnamefont {O.}~\bibnamefont {Brandt}},\ }\href@noop {} {\bibfield
  {journal} {\bibinfo  {journal} {submitted to Phys. Rev. B}\ } (\bibinfo
  {year} {2014}{\natexlab{a}})},\ \Eprint {http://arxiv.org/abs/1408.1236}
  {arXiv:1408.1236} \BibitemShut {NoStop}%
\bibitem [{\citenamefont {Rudolph}\ \emph {et~al.}(2013)\citenamefont
  {Rudolph}, \citenamefont {Schweickert}, \citenamefont {Mork\"{o}tter},
  \citenamefont {Hanschke}, \citenamefont {Hertenberger}, \citenamefont
  {Bichler}, \citenamefont {Koblm\"{u}ller}, \citenamefont {Abstreiter},\ and\
  \citenamefont {Finley}}]{Rudolph2013}%
  \BibitemOpen
  \bibfield  {author} {\bibinfo {author} {\bibfnamefont {D.}~\bibnamefont
  {Rudolph}}, \bibinfo {author} {\bibfnamefont {L.}~\bibnamefont
  {Schweickert}}, \bibinfo {author} {\bibfnamefont {S.}~\bibnamefont
  {Mork\"{o}tter}}, \bibinfo {author} {\bibfnamefont {L.}~\bibnamefont
  {Hanschke}}, \bibinfo {author} {\bibfnamefont {S.}~\bibnamefont
  {Hertenberger}}, \bibinfo {author} {\bibfnamefont {M.}~\bibnamefont
  {Bichler}}, \bibinfo {author} {\bibfnamefont {G.}~\bibnamefont
  {Koblm\"{u}ller}}, \bibinfo {author} {\bibfnamefont {G.}~\bibnamefont
  {Abstreiter}}, \ and\ \bibinfo {author} {\bibfnamefont {J.~J.}\ \bibnamefont
  {Finley}},\ }\href {\doibase 10.1088/1367-2630/15/11/113032} {\bibfield
  {journal} {\bibinfo  {journal} {New J. Phys.}\ }\textbf {\bibinfo {volume}
  {15}},\ \bibinfo {pages} {113032} (\bibinfo {year} {2013})}\BibitemShut
  {NoStop}%
\bibitem [{\citenamefont {Wu}\ \emph {et~al.}(2008)\citenamefont {Wu},
  \citenamefont {Fischer}, \citenamefont {Ponce}, \citenamefont {Bastek},
  \citenamefont {Christen}, \citenamefont {Wernicke}, \citenamefont {Weyers},\
  and\ \citenamefont {Kneissl}}]{Wu2008}%
  \BibitemOpen
  \bibfield  {author} {\bibinfo {author} {\bibfnamefont {Z.~H.}\ \bibnamefont
  {Wu}}, \bibinfo {author} {\bibfnamefont {A.~M.}\ \bibnamefont {Fischer}},
  \bibinfo {author} {\bibfnamefont {F.~A.}\ \bibnamefont {Ponce}}, \bibinfo
  {author} {\bibfnamefont {B.}~\bibnamefont {Bastek}}, \bibinfo {author}
  {\bibfnamefont {J.}~\bibnamefont {Christen}}, \bibinfo {author}
  {\bibfnamefont {T.}~\bibnamefont {Wernicke}}, \bibinfo {author}
  {\bibfnamefont {M.}~\bibnamefont {Weyers}}, \ and\ \bibinfo {author}
  {\bibfnamefont {M.}~\bibnamefont {Kneissl}},\ }\href {\doibase
  http://dx.doi.org/10.1063/1.2918834} {\bibfield  {journal} {\bibinfo
  {journal} {Appl. Phys. Lett.}\ }\textbf {\bibinfo {volume} {92}},\ \bibinfo
  {eid} {171904} (\bibinfo {year} {2008})}\BibitemShut {NoStop}%
\bibitem [{\citenamefont {Hauswald}\ \emph {et~al.}(2013)\citenamefont
  {Hauswald}, \citenamefont {Flissikowski}, \citenamefont {Gotschke},
  \citenamefont {Calarco}, \citenamefont {Geelhaar}, \citenamefont {Grahn},\
  and\ \citenamefont {Brandt}}]{Hauswald2013}%
  \BibitemOpen
  \bibfield  {author} {\bibinfo {author} {\bibfnamefont {C.}~\bibnamefont
  {Hauswald}}, \bibinfo {author} {\bibfnamefont {T.}~\bibnamefont
  {Flissikowski}}, \bibinfo {author} {\bibfnamefont {T.}~\bibnamefont
  {Gotschke}}, \bibinfo {author} {\bibfnamefont {R.}~\bibnamefont {Calarco}},
  \bibinfo {author} {\bibfnamefont {L.}~\bibnamefont {Geelhaar}}, \bibinfo
  {author} {\bibfnamefont {H.~T.}\ \bibnamefont {Grahn}}, \ and\ \bibinfo
  {author} {\bibfnamefont {O.}~\bibnamefont {Brandt}},\ }\href {\doibase
  10.1103/PhysRevB.88.075312} {\bibfield  {journal} {\bibinfo  {journal} {Phys.
  Rev. B}\ }\textbf {\bibinfo {volume} {88}},\ \bibinfo {pages} {075312}
  (\bibinfo {year} {2013})}\BibitemShut {NoStop}%
\bibitem [{\citenamefont {Monemar}\ \emph {et~al.}(2010)\citenamefont
  {Monemar}, \citenamefont {Paskov}, \citenamefont {Bergman}, \citenamefont
  {Pozina}, \citenamefont {Toropov}, \citenamefont {Shubina}, \citenamefont
  {Malinauskas},\ and\ \citenamefont {Usui}}]{Monemar2010}%
  \BibitemOpen
  \bibfield  {author} {\bibinfo {author} {\bibfnamefont {B.}~\bibnamefont
  {Monemar}}, \bibinfo {author} {\bibfnamefont {P.~P.}\ \bibnamefont {Paskov}},
  \bibinfo {author} {\bibfnamefont {J.~P.}\ \bibnamefont {Bergman}}, \bibinfo
  {author} {\bibfnamefont {G.}~\bibnamefont {Pozina}}, \bibinfo {author}
  {\bibfnamefont {A.~A.}\ \bibnamefont {Toropov}}, \bibinfo {author}
  {\bibfnamefont {T.~V.}\ \bibnamefont {Shubina}}, \bibinfo {author}
  {\bibfnamefont {T.}~\bibnamefont {Malinauskas}}, \ and\ \bibinfo {author}
  {\bibfnamefont {A.}~\bibnamefont {Usui}},\ }\href {\doibase
  10.1103/PhysRevB.82.235202} {\bibfield  {journal} {\bibinfo  {journal} {Phys.
  Rev. B}\ }\textbf {\bibinfo {volume} {82}},\ \bibinfo {pages} {235202}
  (\bibinfo {year} {2010})}\BibitemShut {NoStop}%
\bibitem [{\citenamefont {Lahourcade}\ \emph {et~al.}(2008)\citenamefont
  {Lahourcade}, \citenamefont {Renard}, \citenamefont {Gayral}, \citenamefont
  {Monroy}, \citenamefont {Chauvat},\ and\ \citenamefont
  {Ruterana}}]{Lahourcade2008}%
  \BibitemOpen
  \bibfield  {author} {\bibinfo {author} {\bibfnamefont {L.}~\bibnamefont
  {Lahourcade}}, \bibinfo {author} {\bibfnamefont {J.}~\bibnamefont {Renard}},
  \bibinfo {author} {\bibfnamefont {B.}~\bibnamefont {Gayral}}, \bibinfo
  {author} {\bibfnamefont {E.}~\bibnamefont {Monroy}}, \bibinfo {author}
  {\bibfnamefont {M.~P.}\ \bibnamefont {Chauvat}}, \ and\ \bibinfo {author}
  {\bibfnamefont {P.}~\bibnamefont {Ruterana}},\ }\href {\doibase
  10.1063/1.2908205} {\bibfield  {journal} {\bibinfo  {journal} {J. Appl.
  Phys.}\ }\textbf {\bibinfo {volume} {103}},\ \bibinfo {pages} {093514}
  (\bibinfo {year} {2008})}\BibitemShut {NoStop}%
\bibitem [{\citenamefont {Badcock}\ \emph {et~al.}(2012)\citenamefont
  {Badcock}, \citenamefont {Kappers}, \citenamefont {Moram}, \citenamefont
  {Dawson},\ and\ \citenamefont {Humphreys}}]{Badcock2012}%
  \BibitemOpen
  \bibfield  {author} {\bibinfo {author} {\bibfnamefont {T.~J.}\ \bibnamefont
  {Badcock}}, \bibinfo {author} {\bibfnamefont {M.~J.}\ \bibnamefont
  {Kappers}}, \bibinfo {author} {\bibfnamefont {M.~A.}\ \bibnamefont {Moram}},
  \bibinfo {author} {\bibfnamefont {P.}~\bibnamefont {Dawson}}, \ and\ \bibinfo
  {author} {\bibfnamefont {C.~J.}\ \bibnamefont {Humphreys}},\ }\href {\doibase
  10.1002/pssb.201100480} {\bibfield  {journal} {\bibinfo  {journal} {Physica
  Status Solidi B}\ }\textbf {\bibinfo {volume} {249}},\ \bibinfo {pages} {498}
  (\bibinfo {year} {2012})}\BibitemShut {NoStop}%
\bibitem [{\citenamefont {Kagaya}\ \emph {et~al.}(2011)\citenamefont {Kagaya},
  \citenamefont {Corfdir}, \citenamefont {Gani\`{e}re}, \citenamefont
  {Deveaud-Pl\'{e}dran}, \citenamefont {Grandjean},\ and\ \citenamefont
  {Chichibu}}]{Kagaya2011}%
  \BibitemOpen
  \bibfield  {author} {\bibinfo {author} {\bibfnamefont {M.}~\bibnamefont
  {Kagaya}}, \bibinfo {author} {\bibfnamefont {P.}~\bibnamefont {Corfdir}},
  \bibinfo {author} {\bibfnamefont {J.-D.}\ \bibnamefont {Gani\`{e}re}},
  \bibinfo {author} {\bibfnamefont {B.}~\bibnamefont {Deveaud-Pl\'{e}dran}},
  \bibinfo {author} {\bibfnamefont {N.}~\bibnamefont {Grandjean}}, \ and\
  \bibinfo {author} {\bibfnamefont {S.~F.}\ \bibnamefont {Chichibu}},\ }\href
  {\doibase 10.1143/JJAP.50.111002} {\bibfield  {journal} {\bibinfo  {journal}
  {Jpn. J. Appl. Phys.}\ }\textbf {\bibinfo {volume} {50}},\ \bibinfo {pages}
  {111002} (\bibinfo {year} {2011})}\BibitemShut {NoStop}%
\bibitem [{\citenamefont {Furusawa}\ \emph {et~al.}(2013)\citenamefont
  {Furusawa}, \citenamefont {Ishikawa}, \citenamefont {Tashiro}, \citenamefont
  {Hazu}, \citenamefont {Nagao}, \citenamefont {Ikeda}, \citenamefont
  {Fujito},\ and\ \citenamefont {Chichibu}}]{Furusawa2013}%
  \BibitemOpen
  \bibfield  {author} {\bibinfo {author} {\bibfnamefont {K.}~\bibnamefont
  {Furusawa}}, \bibinfo {author} {\bibfnamefont {Y.}~\bibnamefont {Ishikawa}},
  \bibinfo {author} {\bibfnamefont {M.}~\bibnamefont {Tashiro}}, \bibinfo
  {author} {\bibfnamefont {K.}~\bibnamefont {Hazu}}, \bibinfo {author}
  {\bibfnamefont {S.}~\bibnamefont {Nagao}}, \bibinfo {author} {\bibfnamefont
  {H.}~\bibnamefont {Ikeda}}, \bibinfo {author} {\bibfnamefont
  {K.}~\bibnamefont {Fujito}}, \ and\ \bibinfo {author} {\bibfnamefont {S.~F.}\
  \bibnamefont {Chichibu}},\ }\href {\doibase 10.1063/1.4817297} {\bibfield
  {journal} {\bibinfo  {journal} {Appl. Phys. Lett.}\ }\textbf {\bibinfo
  {volume} {103}},\ \bibinfo {pages} {052108} (\bibinfo {year}
  {2013})}\BibitemShut {NoStop}%
\bibitem [{\citenamefont {Corfdir}\ \emph {et~al.}(2012)\citenamefont
  {Corfdir}, \citenamefont {Dussaigne}, \citenamefont {Teisseyre},
  \citenamefont {Suski}, \citenamefont {Grzegory}, \citenamefont {Lefebvre},
  \citenamefont {Giraud}, \citenamefont {Gani\`{e}re}, \citenamefont
  {Grandjean},\ and\ \citenamefont {Deveaud-Pl\'edran}}]{Corfdir2012b}%
  \BibitemOpen
  \bibfield  {author} {\bibinfo {author} {\bibfnamefont {P.}~\bibnamefont
  {Corfdir}}, \bibinfo {author} {\bibfnamefont {A.}~\bibnamefont {Dussaigne}},
  \bibinfo {author} {\bibfnamefont {H.}~\bibnamefont {Teisseyre}}, \bibinfo
  {author} {\bibfnamefont {T.}~\bibnamefont {Suski}}, \bibinfo {author}
  {\bibfnamefont {I.}~\bibnamefont {Grzegory}}, \bibinfo {author}
  {\bibfnamefont {P.}~\bibnamefont {Lefebvre}}, \bibinfo {author}
  {\bibfnamefont {E.}~\bibnamefont {Giraud}}, \bibinfo {author} {\bibfnamefont
  {J.-D.}\ \bibnamefont {Gani\`{e}re}}, \bibinfo {author} {\bibfnamefont
  {N.}~\bibnamefont {Grandjean}}, \ and\ \bibinfo {author} {\bibfnamefont
  {B.}~\bibnamefont {Deveaud-Pl\'edran}},\ }\href {\doibase
  http://dx.doi.org/10.1063/1.3681816} {\bibfield  {journal} {\bibinfo
  {journal} {J. Appl. Phys.}\ }\textbf {\bibinfo {volume} {111}},\ \bibinfo
  {eid} {033517} (\bibinfo {year} {2012})}\BibitemShut {NoStop}%
\bibitem [{\citenamefont {Rashba}\ and\ \citenamefont
  {Gurgenishvili}(1962)}]{Rashba1962}%
  \BibitemOpen
  \bibfield  {author} {\bibinfo {author} {\bibfnamefont {E.~I.}\ \bibnamefont
  {Rashba}}\ and\ \bibinfo {author} {\bibfnamefont {G.~E.}\ \bibnamefont
  {Gurgenishvili}},\ }\href@noop {} {\bibfield  {journal} {\bibinfo  {journal}
  {Sov. Phys. - Solid State}\ }\textbf {\bibinfo {volume} {4}},\ \bibinfo
  {pages} {759} (\bibinfo {year} {1962})}\BibitemShut {NoStop}%
\bibitem [{\citenamefont {'t~Hooft}\ \emph {et~al.}(1987)\citenamefont
  {'t~Hooft}, \citenamefont {van~der Poel}, \citenamefont {Molenkamp},\ and\
  \citenamefont {Foxon}}]{Hooft1987}%
  \BibitemOpen
  \bibfield  {author} {\bibinfo {author} {\bibfnamefont {G.~W.}\ \bibnamefont
  {'t~Hooft}}, \bibinfo {author} {\bibfnamefont {W.~A. J.~A.}\ \bibnamefont
  {van~der Poel}}, \bibinfo {author} {\bibfnamefont {L.~W.}\ \bibnamefont
  {Molenkamp}}, \ and\ \bibinfo {author} {\bibfnamefont {C.~T.}\ \bibnamefont
  {Foxon}},\ }\href {\doibase 10.1103/PhysRevB.35.8281} {\bibfield  {journal}
  {\bibinfo  {journal} {Phys. Rev. B}\ }\textbf {\bibinfo {volume} {35}},\
  \bibinfo {pages} {8281} (\bibinfo {year} {1987})}\BibitemShut {NoStop}%
\bibitem [{\citenamefont {Andreani}\ \emph {et~al.}(1991)\citenamefont
  {Andreani}, \citenamefont {Tassone},\ and\ \citenamefont
  {Bassani}}]{Andreani1991}%
  \BibitemOpen
  \bibfield  {author} {\bibinfo {author} {\bibfnamefont {L.~C.}\ \bibnamefont
  {Andreani}}, \bibinfo {author} {\bibfnamefont {F.}~\bibnamefont {Tassone}}, \
  and\ \bibinfo {author} {\bibfnamefont {F.}~\bibnamefont {Bassani}},\ }\href
  {\doibase 10.1016/0038-1098(91)90761-J} {\bibfield  {journal} {\bibinfo
  {journal} {Solid State Commun.}\ }\textbf {\bibinfo {volume} {77}},\ \bibinfo
  {pages} {641} (\bibinfo {year} {1991})}\BibitemShut {NoStop}%
\bibitem [{\citenamefont {Citrin}(1992)}]{Citrin1993a}%
  \BibitemOpen
  \bibfield  {author} {\bibinfo {author} {\bibfnamefont {D.~S.}\ \bibnamefont
  {Citrin}},\ }\href {\doibase 10.1103/PhysRevLett.69.3393} {\bibfield
  {journal} {\bibinfo  {journal} {Phys. Rev. Lett.}\ }\textbf {\bibinfo
  {volume} {69}},\ \bibinfo {pages} {3393} (\bibinfo {year}
  {1992})}\BibitemShut {NoStop}%
\bibitem [{\citenamefont {Dussaigne}\ \emph {et~al.}(2011)\citenamefont
  {Dussaigne}, \citenamefont {Corfdir}, \citenamefont {Levrat}, \citenamefont
  {Zhu}, \citenamefont {Martin}, \citenamefont {Lefebvre}, \citenamefont
  {Gani\`{e}re}, \citenamefont {Butt\'{e}}, \citenamefont
  {Deveaud-Pl\'{e}dran}, \citenamefont {Grandjean}, \citenamefont {Arroyo},\
  and\ \citenamefont {Stadelmann}}]{Dussaigne2011}%
  \BibitemOpen
  \bibfield  {author} {\bibinfo {author} {\bibfnamefont {A.}~\bibnamefont
  {Dussaigne}}, \bibinfo {author} {\bibfnamefont {P.}~\bibnamefont {Corfdir}},
  \bibinfo {author} {\bibfnamefont {J.}~\bibnamefont {Levrat}}, \bibinfo
  {author} {\bibfnamefont {T.}~\bibnamefont {Zhu}}, \bibinfo {author}
  {\bibfnamefont {D.}~\bibnamefont {Martin}}, \bibinfo {author} {\bibfnamefont
  {P.}~\bibnamefont {Lefebvre}}, \bibinfo {author} {\bibfnamefont {J.-D.}\
  \bibnamefont {Gani\`{e}re}}, \bibinfo {author} {\bibfnamefont
  {R.}~\bibnamefont {Butt\'{e}}}, \bibinfo {author} {\bibfnamefont
  {B.}~\bibnamefont {Deveaud-Pl\'{e}dran}}, \bibinfo {author} {\bibfnamefont
  {N.}~\bibnamefont {Grandjean}}, \bibinfo {author} {\bibfnamefont
  {Y.}~\bibnamefont {Arroyo}}, \ and\ \bibinfo {author} {\bibfnamefont
  {P.}~\bibnamefont {Stadelmann}},\ }\href {\doibase
  10.1088/0268-1242/26/2/025012} {\bibfield  {journal} {\bibinfo  {journal}
  {Semicond. Sci. Technol.}\ }\textbf {\bibinfo {volume} {26}},\ \bibinfo
  {pages} {025012} (\bibinfo {year} {2011})}\BibitemShut {NoStop}%
\bibitem [{\citenamefont {Citrin}(1993)}]{Citrin1993}%
  \BibitemOpen
  \bibfield  {author} {\bibinfo {author} {\bibfnamefont {D.~S.}\ \bibnamefont
  {Citrin}},\ }\href {\doibase 10.1103/PhysRevB.47.3832} {\bibfield  {journal}
  {\bibinfo  {journal} {Phys. Rev. B}\ }\textbf {\bibinfo {volume} {47}},\
  \bibinfo {pages} {3832} (\bibinfo {year} {1993})}\BibitemShut {NoStop}%
\bibitem [{\citenamefont {Rosales}\ \emph {et~al.}(2014)\citenamefont
  {Rosales}, \citenamefont {Gil}, \citenamefont {Bretagnon}, \citenamefont
  {Guizal}, \citenamefont {Zhang}, \citenamefont {Okur}, \citenamefont
  {Monavarian}, \citenamefont {Izyumskaya}, \citenamefont {Avrutin},
  \citenamefont {Ozgur}, \citenamefont {Morko\c{c}},\ and\ \citenamefont
  {Leach}}]{Rosales2014}%
  \BibitemOpen
  \bibfield  {author} {\bibinfo {author} {\bibfnamefont {D.}~\bibnamefont
  {Rosales}}, \bibinfo {author} {\bibfnamefont {B.}~\bibnamefont {Gil}},
  \bibinfo {author} {\bibfnamefont {T.}~\bibnamefont {Bretagnon}}, \bibinfo
  {author} {\bibfnamefont {B.}~\bibnamefont {Guizal}}, \bibinfo {author}
  {\bibfnamefont {F.}~\bibnamefont {Zhang}}, \bibinfo {author} {\bibfnamefont
  {S.}~\bibnamefont {Okur}}, \bibinfo {author} {\bibfnamefont {M.}~\bibnamefont
  {Monavarian}}, \bibinfo {author} {\bibfnamefont {N.}~\bibnamefont
  {Izyumskaya}}, \bibinfo {author} {\bibfnamefont {V.}~\bibnamefont {Avrutin}},
  \bibinfo {author} {\bibfnamefont {U.}~\bibnamefont {Ozgur}}, \bibinfo
  {author} {\bibfnamefont {H.}~\bibnamefont {Morko\c{c}}}, \ and\ \bibinfo
  {author} {\bibfnamefont {J.~H.}\ \bibnamefont {Leach}},\ }\href {\doibase
  http://dx.doi.org/10.1063/1.4865959} {\bibfield  {journal} {\bibinfo
  {journal} {J. Appl. Phys.}\ }\textbf {\bibinfo {volume} {115}},\ \bibinfo
  {pages} {073510} (\bibinfo {year} {2014})}\BibitemShut {NoStop}%
\bibitem [{\citenamefont {Sun}\ \emph {et~al.}(2002)\citenamefont {Sun},
  \citenamefont {Brandt}, \citenamefont {Jahn}, \citenamefont {Liu},
  \citenamefont {Trampert}, \citenamefont {Cronenberg}, \citenamefont {Dhar},\
  and\ \citenamefont {Ploog}}]{Sun2002}%
  \BibitemOpen
  \bibfield  {author} {\bibinfo {author} {\bibfnamefont {Y.~J.}\ \bibnamefont
  {Sun}}, \bibinfo {author} {\bibfnamefont {O.}~\bibnamefont {Brandt}},
  \bibinfo {author} {\bibfnamefont {U.}~\bibnamefont {Jahn}}, \bibinfo {author}
  {\bibfnamefont {T.~Y.}\ \bibnamefont {Liu}}, \bibinfo {author} {\bibfnamefont
  {A.}~\bibnamefont {Trampert}}, \bibinfo {author} {\bibfnamefont
  {S.}~\bibnamefont {Cronenberg}}, \bibinfo {author} {\bibfnamefont
  {S.}~\bibnamefont {Dhar}}, \ and\ \bibinfo {author} {\bibfnamefont {K.~H.}\
  \bibnamefont {Ploog}},\ }\href {\doibase 10.1063/1.1513874} {\bibfield
  {journal} {\bibinfo  {journal} {J. Appl. Phys.}\ }\textbf {\bibinfo {volume}
  {92}},\ \bibinfo {pages} {5714} (\bibinfo {year} {2002})}\BibitemShut
  {NoStop}%
\bibitem [{\citenamefont {Gorgis}\ \emph {et~al.}(2012)\citenamefont {Gorgis},
  \citenamefont {Flissikowski}, \citenamefont {Brandt}, \citenamefont
  {Ch\`{e}ze}, \citenamefont {Geelhaar}, \citenamefont {Riechert},\ and\
  \citenamefont {Grahn}}]{Gorgis2012}%
  \BibitemOpen
  \bibfield  {author} {\bibinfo {author} {\bibfnamefont {A.}~\bibnamefont
  {Gorgis}}, \bibinfo {author} {\bibfnamefont {T.}~\bibnamefont
  {Flissikowski}}, \bibinfo {author} {\bibfnamefont {O.}~\bibnamefont
  {Brandt}}, \bibinfo {author} {\bibfnamefont {C.}~\bibnamefont {Ch\`{e}ze}},
  \bibinfo {author} {\bibfnamefont {L.}~\bibnamefont {Geelhaar}}, \bibinfo
  {author} {\bibfnamefont {H.}~\bibnamefont {Riechert}}, \ and\ \bibinfo
  {author} {\bibfnamefont {H.~T.}\ \bibnamefont {Grahn}},\ }\href {\doibase
  10.1103/PhysRevB.86.041302} {\bibfield  {journal} {\bibinfo  {journal} {Phys.
  Rev. B}\ }\textbf {\bibinfo {volume} {86}},\ \bibinfo {pages} {041302(R)}
  (\bibinfo {year} {2012})}\BibitemShut {NoStop}%
\bibitem [{\citenamefont {Schlager}\ \emph {et~al.}(2008)\citenamefont
  {Schlager}, \citenamefont {Bertness}, \citenamefont {Blanchard},
  \citenamefont {Robins}, \citenamefont {Roshko},\ and\ \citenamefont
  {Sanford}}]{Schlager2008}%
  \BibitemOpen
  \bibfield  {author} {\bibinfo {author} {\bibfnamefont {J.~B.}\ \bibnamefont
  {Schlager}}, \bibinfo {author} {\bibfnamefont {K.~A.}\ \bibnamefont
  {Bertness}}, \bibinfo {author} {\bibfnamefont {P.~T.}\ \bibnamefont
  {Blanchard}}, \bibinfo {author} {\bibfnamefont {L.~H.}\ \bibnamefont
  {Robins}}, \bibinfo {author} {\bibfnamefont {A.}~\bibnamefont {Roshko}}, \
  and\ \bibinfo {author} {\bibfnamefont {N.~A.}\ \bibnamefont {Sanford}},\
  }\href {\doibase 10.1063/1.2940732} {\bibfield  {journal} {\bibinfo
  {journal} {J. Appl. Phys.}\ }\textbf {\bibinfo {volume} {103}},\ \bibinfo
  {pages} {124309} (\bibinfo {year} {2008})}\BibitemShut {NoStop}%
\bibitem [{\citenamefont {Hauswald}\ \emph
  {et~al.}(2014{\natexlab{b}})\citenamefont {Hauswald}, \citenamefont
  {Flissikowski}, \citenamefont {Grahn}, \citenamefont {Geelhaar},
  \citenamefont {Riechert},\ and\ \citenamefont {Brandt}}]{Hauswald2014a}%
  \BibitemOpen
  \bibfield  {author} {\bibinfo {author} {\bibfnamefont {C.}~\bibnamefont
  {Hauswald}}, \bibinfo {author} {\bibfnamefont {T.}~\bibnamefont
  {Flissikowski}}, \bibinfo {author} {\bibfnamefont {H.~T.}\ \bibnamefont
  {Grahn}}, \bibinfo {author} {\bibfnamefont {L.}~\bibnamefont {Geelhaar}},
  \bibinfo {author} {\bibfnamefont {H.}~\bibnamefont {Riechert}}, \ and\
  \bibinfo {author} {\bibfnamefont {O.}~\bibnamefont {Brandt}},\ }\href
  {\doibase 10.1117/12.2039082} {\bibfield  {journal} {\bibinfo  {journal}
  {Proc. SPIE}\ }\textbf {\bibinfo {volume} {8986}},\ \bibinfo {pages} {89860V}
  (\bibinfo {year} {2014}{\natexlab{b}})}\BibitemShut {NoStop}%
\end{thebibliography}
\end{document}